\documentclass[12pt]{article}

\usepackage{latexsym}
\usepackage{amssymb,amsfonts,amsmath}
\usepackage{graphicx} 
\usepackage{indentfirst}
\usepackage{bbm}
\usepackage{amssymb}
\usepackage{verbatim}
\usepackage{amsmath, amsthm,amssymb}
\usepackage{mathrsfs}
\usepackage{hyperref}
\usepackage{amsfonts}
\usepackage{dsfont}
\usepackage{cite}
\usepackage{xcolor}
\usepackage[multiple]{footmisc}
\parskip=\medskipamount

\newcommand {\cD}{{\cal D}}
\newcommand {\cE}{{\cal E}}

\newcommand {\cH}{{\cal H}}

\newcommand {\cJ}{{\cal J}}
\newcommand {\cK}{{\cal K}}
\newcommand {\cL}{{\cal L}}
\newcommand {\cM}{{\cal M}}
\newcommand {\cN}{{\cal N}}

\def\a{\alpha}

\def\b{\beta}
\def\c{\chi}
\def\d{\delta}
\def\e{\epsilon}
\def\f{\phi}
\def\g{\gamma}
\def\G{\Gamma}

\def\j{\psi}
\def\k{\kappa}
\def\l{\lambda}
\def\m{\mu}
\def\n{\nu}
\def\o{\omega}

\def\q{\theta}
\def\r{\rho}
\def\s{\sigma}
\def\t{\tau}
\def\u{\upsilon}
\def\x{\xi}
\def\z{\zeta}
\def\D{\Delta}
\def\F{\Phi}
\def\J{\Psi}
\def\L{\Lambda}
\def\O{\Omega}
\def\P{\Pi}

\def\S{\Sigma}
\def\U{\Upsilon}
\def\X{\Xi}

\def\rd{{\rm d}}
\def\ri{{\rm i}}
\def\re{{\rm e}}

\newcommand{\ad}{{\dot{\alpha}}}                           %new
\newcommand{\bd}{{\dot{\beta}}}                            %new
\newcommand{\ve}{\varepsilon}                            %new
\newcommand{\cDB}{{\bar\cD}}                            %new

\newcommand{\pa}{\partial}                           %new
\newcommand{\hf}{\frac12}
\newcommand{\vf}{\varphi}
\newcommand{\be}{\begin{equation}}
\newcommand{\ee}{\end{equation}}
\newcommand{\bea}{\begin{eqnarray}}
\newcommand{\eea}{\end{eqnarray}}
\newcommand{\non}{\nonumber}
\newcommand{\bm}[1]{\mbox{\boldmath$#1$}}

\def\double #1{#1{\hbox{\kern-2pt $#1$}}}

\newcommand{\gd}{{\dot\g}}
\newcommand{\dd}{{\dot\d}}

\newcommand{\ts}{{\tilde{\s}}}

\newcommand{\qb}{{\bar{\theta}}}

\newif\ifdtup

\newcommand{\bsubeq}{\begin{subequations}}
\newcommand{\esubeq}{\end{subequations}}
\numberwithin{equation}{section}

\newcommand{\sSU}{\mathsf{SU}}
\newcommand{\sSL}{\mathsf{SL}}

\newcommand{\sSO}{\mathsf{SO}}
\newcommand{\sU}{\mathsf{U}}

\begin{document}

\begin{titlepage}
\begin{flushright}
December, 2019 \\
\end{flushright}
\vspace{5mm}

\begin{center}
{\Large \bf 
Symmetries of supergravity backgrounds and supersymmetric  field theory}
\end{center}

\begin{center}

{\bf Sergei M. Kuzenko and Emmanouil S. N. Raptakis} \\
\vspace{5mm}

\footnotesize{
{\it Department of Physics M013, The University of Western Australia\\
35 Stirling Highway, Perth W.A. 6009, Australia}}  
~\\
\vspace{2mm}
~\\
Email: \texttt{ 
sergei.kuzenko@uwa.edu.au, emmanouil.raptakis@research.uwa.edu.au}\\
\vspace{2mm}

\end{center}

\begin{abstract}
\baselineskip=14pt
In four spacetime dimensions, all ${\cal N} =1$ supergravity-matter systems  can be formulated in the so-called $\mathsf{U}(1)$ superspace proposed by Howe in 1981. This paper is devoted to the study of those geometric structures which characterise a background $\mathsf{U}(1)$ superspace and are important in the context of supersymmetric field theory in curved space. We introduce (conformal) Killing tensor superfields $\ell_{(\alpha_1 \dots \alpha_m) ({\dot \alpha}_1 \dots {\dot \alpha}_n)}$, with $m$ and $n$ non-negative integers, $m+n>0$, and elaborate on their significance in the following  cases: (i) $m=n=1$; (ii) $m-1=n=0$; and (iii) $m=n>1$. The (conformal) Killing vector superfields $\ell_{\alpha \dot \alpha}$ generate the (conformal) isometries of curved superspace, which are symmetries of every (conformal) supersymmetric field theory. The (conformal) Killing spinor superfields $\ell_{\alpha }$ generate extended (conformal) supersymmetry transformations. The (conformal) Killing tensor superfields with $m=n>1$ prove to generate all higher symmetries of the (massless) massive Wess-Zumino operator.
\end{abstract}
\vspace{5mm}

\vfill

\vfill
\end{titlepage}

\newpage
\renewcommand{\thefootnote}{\arabic{footnote}}
\setcounter{footnote}{0}

\tableofcontents{}
\vspace{1cm}
\bigskip\hrule

\allowdisplaybreaks

%%%%%%%%%%%%%%%%%%%%%%%%%%%%%%%%
%%%%%%%%%%%%%%%%%%%%%%%%%%%%%%%%

\section{Introduction}

In order to construct and study supersymmetric field theories in the presence of background supergravity  fields,  a formalism is required to determine (conformal) isometries of the corresponding curved superspace.\footnote{An important example of a curved superspace is 
 the four-dimensional (4D) $\cN=1$ anti-de Sitter (AdS) superspace 
 \cite{Zumino77,IS}, ${\rm AdS}^{4|4}$.} 
 Such a formalism was developed long ago \cite{BK} within the framework of 
the Grimm-Wess-Zumino (GWZ) geometry \cite{GWZ}, 
which underlies the Wess-Zumino (WZ) formulation for
old minimal supergravity \cite{WZ} (see \cite{WB} for a review)
discovered independently in \cite{Siegel77,old1,old2}.
The key outcomes of the analysis given
in \cite{BK}
may be summarised as follows:
\begin{itemize} 
\item
Rigid symmetries of every superconformal field theory on a curved superspace $\cM^{4|4}$ are 
generated by conformal Killing supervector fields on $\cM^{4|4}$,
$\x^A = (\x^a, \x^\a , \bar \x_\ad)$,  with $\bar \x^a = \x^a$. 
The defining property of $\x^A$  is that the first-order operator $\x^A \cD_A$ 
maps the space of covariantly chiral scalars into itself,
\bea
\bar \cD_\bd \f = 0 \quad \implies \quad \bar \cD_\bd \big(\x^A\cD_A   \f \big) = 0~,
\eea
where $\cD_A= (\cD_a, \cD_\a , \bar \cD^\ad )$ are the superspace covariant derivatives. 
These conditions imply that the spinor component $\x^\a$ is determined in terms of the vector component $\x^a$
as  ${\xi}^{\a}  =  - \frac{ \rm i}{8} \cDB_{\bd} \xi^{\a \bd} $, and the latter  obeys the superconformal Killing equation
\bea
 \cD_{(\b} \xi_{\a) \ad} =0 \quad \Longleftrightarrow \quad
 \bar \cD_{(\bd} \x_{\a\ad)} =0~.
 \label{1.2}
 \eea

\item 
Rigid symmetries of every supersymmetric field theory on $\cM^{4|4}$ are associated with 
those conformal Killing supervector fields $\x^A $ which preserve the volume of 
 the chiral subspace of $\cM^{4|4}$. This condition is equivalent to  
 \bea
 \cD_\a \bar \cD_\ad \x^{\a\ad} = 4G_{\a\ad} \x^{\a\ad}  \quad \implies \quad 
 \cD_a \x^a=0~,
 \label{1.3}
 \eea
 where $G_{\a\ad}$ is the superspace analogue of the  Ricci tensor.
\end{itemize}
Every solution of the equations \eqref{1.2} and \eqref{1.3} 
is called a Killing supervector field.

If $\cM^{4|4}$ is chosen to be Minkowski superspace,  
the general solution of the equation \eqref{1.2} 
corresponds to the ordinary superconformal transformations which span 
$\sSU(2,2|1)$ \cite{Sohnius,Lang,BPT,Shizuya}. In the case of supersymmetric 
curved backgrounds in old minimal supergravity, the equations \eqref{1.2} and \eqref{1.3} 
allow one to obtain all the results described in an influential work 
of Festuccia and Seiberg \cite{FS} and related publications (see e.g. \cite{JS,ST}) 
in the component setting,  as was demonstrated in \cite{K13} 
(see also \cite{K15Corfu} for a review).

 The approach presented in \cite{BK} is  universal, for  
in principle it may  be generalised to curved backgrounds associated with 
any supergravity theory formulated in superspace, see the discussion in \cite{K15Corfu}.
In particular, it has been properly generalised to study supersymmetric backgrounds
in 3D $\cN=2$ supergravity \cite{KLRST-M}, 4D $\cN=2$ supergravity \cite{BIL},
5D $\cN=1$ supergravity \cite{KNT-M} and 6D $\cN=(1,0)$ supergravity \cite{KLT-M_un}. 
It should also be mentioned that this approach 
has been used to construct general rigid supersymmetric field theories
in 5D $\cN=1$ \cite{KT-M07}, 4D $\cN=2$ \cite{KT-M08,BKads,BKLT-M}
and 3D $(p,q)$   \cite{KT-M11,KLT-M12,BKT-M} anti-de Sitter superspaces.

The present paper is aimed, in part, at extending the analysis given in section 6.4 
of \cite{BK} to the so-called $\sU(1)$ superspace geometry proposed by Howe 
in 1981 \cite{Howe} and soon after reviewed 
and further developed in \cite{GGRS}.\footnote{One of the most important original developments presented in \cite{GGRS} is the complete solution of the torsion constraints, which characterise the $\sU(1)$ superspace geometry, in terms of unconstrained superfield prepotentials.}
It is called `$\sU(1)$ superspace' since its structure group 
$\sSL(2,{\mathbb C}) \times \sU(1)_R$ contains the $R$-symmetry factor $\sU(1)_R$ 
that is absent in the case of the GWZ geometry \cite{GWZ}.
The $\sU(1)$ superspace is a powerful setting to formulate $\cN=1$ supergravity-matter systems for two reasons. Firstly, it allows us to describe conformal supergravity 
by including the super-Weyl transformations in the supergravity gauge group. 
Secondly, every off-shell formulation for $\cN=1$ supergravity can be realised as a super-Weyl invariant coupling of conformal supergravity to a compensating supermultiplet $\X$.
In fact, similar properties also hold in the case of the GWZ geometry. 
One may then ask a natural question: What is the point of introducing  $\sU(1)$ 
superspace if  the GWZ geometry allows one to achieve the same goals?
There are at least three answers to this question. Firstly, 
the GWZ geometry is a gauge-fixed version of $\sU(1)$ superspace 
in the sense that the former is obtained from the latter by partially fixing the super-Weyl 
gauge symmetry. Secondly, since the super-Weyl and local $\sU(1)_R$ 
transformations are described by unconstrained real parameters in $\sU(1)$ superspace, these local symmetries may be used to gauge away any compensating scalar supermultiplet $\X$
by imposing the  condition $\X=1$. In the case of the GWZ geometry, 
such a gauge fixing is possible only in the case of old minimal supergravity.
Thirdly, $\sU(1)$ superspace is more useful for describing the new minimal formulation 
of $\cN=1$ supergravity \cite{new}.\footnote{There exists 
an alternative formulation for conformal supergravity,  the so-called conformal superspace approach
\cite{Butter4DN=1}, which is more general  than $\sU(1)$ superspace in the sense that
the latter is obtained from the former by partially fixing the gauge freedom. 
When studying the symmetries of supergravity backgrounds, however, 
$\sU(1)$ superspace is more economical for applications to deal with. 
}

Along with the (conformal) Killing vector superfields $\x_{\a\ad}$, which generate 
the (conformal) isometries of a curved superspace $\cM^{4|4}$, 
in this paper (Sections \ref{section4} and \ref{section5})
we will analyse the structure of (conformal) Killing tensor superfields 
$\ell_{\a(m) \ad(n)} =\ell_{(\a_1 \dots \a_m) (\ad_1 \dots \ad_n)}$,
with $m$ and $n$ non-negative integers, $m+n>0$.
Some of the motivations to study these supersymmetric extensions of the (conformal)
Killing tensor fields are similar to those that have been pursued in the non-supersymmetric case, which are:
(i) higher-order integrals of motion, see e.g. \cite{WP}; 
(ii) new conserved currents from old ones, see e.g. \cite{Mikhailov}; and 
(iii) higher symmetries of relativistic wave equations, see e.g.
\cite{Nikitin,NikitinP,ShSh,ShV,Eastwood,Vasiliev2004}. There are also conceptually new 
motivations. In particular, if a curved superspace $\cM^{4|4}$ possesses a 
(conformal) Killing spinor superfield $\ell_\a$, extended supersymmetric field theories 
may be constructed, including superconformal nonlinear  $\s$-models 
on hyperk\"ahler cones, see section \ref{section4.4}.

The concept of a Killing tensor superfield $\ell_{\a(n) \ad(n)} = \bar \ell_{\a(n) \ad(n)} $ 
was introduced in 1997 \cite{GKS} in the framework of $\cN=1$ AdS supersymmetry.
There are two types of constraints obeyed by $\ell_{\a(n) \ad(n)} $, which are:
\begin{subequations}\label{1.4}
\bea
\cD_{(\a_1} \ell_{\a_2 \dots \a_{n+1} ) \ad(n)} &=& 0 \quad \Longleftrightarrow  \quad 
\bar \cD_{(\ad_1} \ell_{\a(n) \ad_2 \dots \ad_{n+1} ) } = 0 ~, \label{1.4a}\\
\cD^\b \bar \cD^\bd \ell_{\b \a_1 \dots \a_{n-1} \bd \ad_1 \dots \ad_{n-1}} &=&0
\quad \Longleftrightarrow  \quad 
\bar \cD^\bd  \cD^\b \ell_{\b \a_1 \dots \a_{n-1} \bd \ad_1 \dots \ad_{n-1}} =0~.
\label{1.4b}
\eea
\end{subequations}
These differential constraints have a natural origin in the context of 
the two dually equivalent gauge models for the massless superspin-$(n+\hf)$ multiplet in AdS${}^{4|4}$ which were proposed in \cite{KS94}. The dynamical variables of these models 
consist of a gauge superfield and a compensating supermultiplet.
In both models the gauge superfield is the same, that is 
a real unconstrained superconformal prepotential $H_{\a(n) \ad(n)}$, 
while the compensators are different.
 In one model the compensator is a transverse linear superfield
$\G_{\a(n-2) \ad(n-2)}$, and in the other is it  a longitudinal linear superfield $G_{\a(n-2) \ad(n-2)}$.\footnote{The terminology follows \cite{KS94,KSP}.} 
The corresponding constraints are 
\begin{subequations} 
\bea
 \bar \cD^\bd \G_{ \a(n-2) \bd \ad(n - 3) } &=& 0 ~,  
\\
 \bar \cD_{(\ad_1} G_{\a(n-2)\ad_2 \dots \ad_{n-1} )} &=& 0  ~.
\eea
\end{subequations}
Equation \eqref{1.4a} 
means that the gauge variation of  
 $H_{\a(n) \ad(n)}$ is equal to zero if the gauge parameter is chosen to be  $\ell_{\a(n) \ad(n)}$. In addition, requiring the gauge variation of the compensator
 (either the transverse or the longitudinal one)  to vanish 
 leads to the equation \eqref{1.4b}. 
It was shown in \cite{GKS} that the space of Killing tensor superfields  $\ell_{\a(n) \ad(n)}$
can be endowed with the structure of a superalgebra, which is one of the 
 higher-spin superalgebras  constructed by Fradkin and Vasiliev \cite{FV1,FV2,Vasiliev88} (see also \cite{KV1,KV2}),
 with respect to the bracket \eqref{4.56} restricted to AdS${}^{4|4}$.
A conformal Killing tensor superfield  $\ell_{\a(n) \ad(n)} $ in AdS${}^{4|4}$ is obtained 
by removing the condition \eqref{1.4b} which is not compatible with the superconformal symmetry 
(this aspect was not discussed explicitly in \cite{GKS}).

In 2016, Howe and Lindstr\"om \cite{HL1} (see also \cite{HL2}) generalised the notion of a conformal Killing tensor to superspace. In the case of $\cN=1$ AdS supersymmetry, 
their definition is equivalent to imposing the condition \eqref{1.4a}.
Our definition of conformal Killing tensor superfields in curved superspace 
differs from the one given in \cite{HL1}, however they prove to be equivalent. 

This paper is organised as follows. Section 2 is devoted to a brief review of  $\sU(1)$ superspace.
The conformal isometries of a supergravity background are studied in Section 3. 
We also describe the action principle for superconformal field theories in a curved superspace and give an example of such dynamical systems -- a superconformal nonlinear $\s$-model.
Section 4 is devoted to a systematic study of conformal Killing tensor superfields $\ell_{\a(m)\ad(n)}$ in curved superspace. We demonstrate the significance of different types of conformal Killing tensor superfields for various superconformal field theories 
in curved superspace. 
The isometries of a supergravity background are studied in Section 5. 
We also introduce Killing spinor $\ell_\a$ and tensor $\ell_{\a(n) \ad(n)} $ superfields
and demonstrate their significance for several supersymmetric field theories 
in curved superspace. The symmetries of bosonic supergravity backgrounds are 
studied in Section 6. Concluding comments are given in Section 7. 
The main body of the paper is accompanied by several technical appendices. 
Appendix A is devoted to the closed super 4-form which describes the chiral action principle. Appendix B concerns various aspects of the component reduction.
The Weyl multiplet gauge is introduced in Appendix C. 

%%%%%%%%%%%%%%%%%%%%%%%%%%%
%%%%%%%%%%%%%%%%%%%%%%%%%%%%

\section{The ABC of $\sU(1)$ superspace}

In this section we review the structure of  $\sU(1)$ superspace \cite{Howe,GGRS}.
Our presentation is analogous to \cite{ButterK}.

\subsection{The geometry of $\sU(1)$ superspace}

We consider a curved $\cN=1$ superspace $\mathcal{M}^{4|4}$
parametrised by local coordinates 
$z^{M} = 
(x^{m},\theta^{\m},\bar \theta_{\dot{\mu}})$.  
Its
structure group is chosen to be $\sSL \left(2 , \mathbb{C} \right) \times \sU(1)_{R}$ 
and so the covariant derivatives $\cD_{A} = \left( \cD_{a}, \cD_{\a}, \cDB^{\ad} \right)$ 
have the form
\bea
\cD_{A} = E_{A} + \O_{A} + {\rm i}\, \F_{A} \mathbb{A} ~.
\label{2.1}
\eea
Here $E_A$ denotes the frame field, 
$E_{A}= E_{A}{}^{M} \partial_{M}$, 
with $E_A{}^M$ being the inverse vielbein.
The Lorentz connection  $\O_{A}$ can be written in two different forms,
\bea
\O_{A} = \frac{1}{2} \O_{A}{}^{bc} M_{bc} = \O_{A}{}^{\b \g} M_{\b \g} + \bar \O_{A}{}^{\bd \gd} \bar M_{\bd \gd} ~,
\eea
depending on whether the Lorentz generators with vector ($M_{bc}=-M_{cb}$)
or spinor $( M_{\b\g}=M_{\g\b}$ and $ {\bar M}_{\bd \gd} = \bar M_{\gd\bd})$ indices are used. 
The Lorentz generators act on vectors and Weyl spinors as follows:
\bea
M_{ab} V_{c} = 2 \eta_{c[a} V_{b]} ~, \qquad 
M_{\a \b} \j_{\g} = \ve_{\g (\a} \j_{\b)} ~, \qquad \bar{M}_{\ad \bd} \bar \j_{\gd} = \ve_{\gd ( \ad} \bar \j_{\bd )} ~.
\eea
The last term in \eqref{2.1} is the $\sU(1)_{R}$ connection,
with  the  $R$-symmetry generator  $\mathbb{A}$
being normalised by
\bea
[ \mathbb{A} , \cD_{\a} ] = - \cD_{\a} ~, \qquad [ \mathbb{A} , \cDB_{\ad} ] = + \cDB_{\ad} ~. 
\eea

The supergravity gauge freedom includes local $\cK$-transformations of the form 
\bea
\label{gaugeTf}
\delta_{\mathcal{K}} \cD_{A} = \left[ \mathcal{K} , \cD_{A} \right] ~, 
\qquad \mathcal{K} = \xi^{B} \cD_{B} + K^{\b\g} M_{ \b \g} 
+ \bar K^{\bd \gd} \bar M_{\bd \gd} + {\rm i} \,\r \mathbb{A} ~.
\eea
Here the gauge parameter $\mathcal{K}$ incorporates several parameters describing 
the general coordinate  ($\xi^{B}$), local Lorentz ($K^{ \b \g}$ and $\bar K^{\bd \gd}$)
 and local chiral ($\r$) transformations.
Given a tensor superfield $U$ (with suppressed indices),
its $\cK$-transformation law is
\bea
\label{tensorgaugeTf}
\delta_{\mathcal{K}} U = \mathcal{K} U ~.
\eea

The covariant derivatives obey graded commutation relations 
\bea
[ \cD_{A} , \cD_{B} \} = \mathcal{T}_{AB}{}^{C} \cD_{C} + \mathcal{R}_{AB}{}^{\g \d} M_{\g \d} + \bar{\mathcal{R}}_{AB}{}^{\gd \dd} \bar M_{\gd \dd} + {\rm i} \,\mathcal{F}_{AB} \mathbb{A} ~,
\eea
where $\mathcal{T}_{AB}{}^{C}$ is the torsion, $\mathcal{R}_{AB}{}^{\g \d}$ and its conjugate $\mathcal{R}_{AB}{}^{\gd \dd} $ constitute 
the Lorentz curvature, and $\mathcal{F}_{AB}$ is the $\sU(1)_{R}$ field strength. 
To describe conformal supergravity, 
the covariant derivatives have to obey certain constraints \cite{Howe}.
Their solution is given by the relations
\begin{subequations}
\label{algebra}
\bea
\{ \cD_{\a}, \cD_{\b} \} &=& -4{\bar R} M_{\a \b}~, \qquad
\{\cDB_{\ad}, \cDB_{\bd} \} =  4R {\bar M}_{\ad \bd}~, \\
&& {} \qquad \{ \cD_{\a} , \cDB_{\ad} \} = -2{\rm i} \cD_{\a \ad} ~, 
 \\
\left[ \cD_{\a} , \cD_{ \b \bd } \right]
     & = &
     {\rm i}
     {\ve}_{\a \b}
\Big({\bar R}\,\cDB_\bd + G^\g{}_\bd \cD_\g
- (\cD^\g G^\d{}_\bd)  M_{\g \d}
+2{\bar W}_\bd{}^{\gd \dot{\d}}
{\bar M}_{\gd \dot{\d} }  \Big) \non \\
&&
+ {\rm i} (\cDB_{\bd} {\bar R})  M_{\a \b}
-\frac{\ri}{3} \ve_{\a\b} \bar X^\gd \bar M_{\gd \bd} + \frac{\ri}{2} \ve_{\a\b} \bar X_\bd {\mathbb A}
~, \\
\left[ {\bar \cD}_{\ad} , \cD_{\b\bd} \right]
     & = &
     - {\rm i}
     \ve_{\ad\bd}
\Big({R}\,\cD_{\b} + G_\b{}^\gd \cDB_\gd
- (\cDB^{\gd} G_{\b}{}^{\dd})  \bar M_{\gd \dd}
+2{W}_\b{}^{\g \d}
{M}_{\g \d }  \Big) \non \\
&&
- {\rm i} (\cD_\b R)  {\bar M}_{\ad \bd}
+\frac{\ri}{3} \ve_{\ad \bd} X^{\g} M_{\g \b} + \frac{\ri}{2} \ve_{\ad\bd} X_\b {\mathbb A}
~,
\eea
which lead to 
\bea
\label{vectorCommutator}
\left[ \cD_{\a \ad} , \cD_{\b \bd} \right] & = & \ve_{\a \b} \bar \psi_{\ad \bd} + \ve_{\ad \bd} \psi_{\a \b} ~, \\
\psi_{\a \b} & = & - \ri G_{ ( \a }{}^{\gd} \cD_{\b ) \gd} + \frac{1}{2} \cD_{( \a } R \cD_{\b)} + \frac{1}{2} \cD_{ ( \a } G_{\b )}{}^{\gd} \cDB_{\gd} + W_{\a \b}{}^{\g} \cD_{\g} \non \\
&& + \frac{1}{6} X_{( \a} \cD_{ \b)} + \frac{1}{4} (\cD^{2} - 8R) {\bar R} M_{\a \b} + \cD_{( \a} W_{ \b)}{}^{\g \d} M_{\g \d} \non \\
&& - \frac{1}{6} \cD_{( \a} X^{\g} M_{\b) \g} - \frac{1}{2} \cD_{ ( \a} \cDB^{\gd} G_{\b)}{}^{\dd} {\bar M}_{\gd \dd} - \frac{1}{4} \cD_{( \a} X_{\b)} \mathbb{A} ~, \\
{\bar \psi}_{\ad \bd} & = & \ri G^{\g}{}_{( \ad} \cD_{\g \bd)} - \frac{1}{2} \cDB_{( \ad } {\bar R} \cDB_{\bd)} - \frac{1}{2} \cDB_{ ( \ad } G^{\g}{}_{\bd)} \cD_{\g} - {\bar W}_{\ad \bd}{}^{\gd} \cDB_{\gd} \non \\
&& - \frac{1}{6} {\bar X}_{( \ad} \cDB_{ \bd)} + \frac{1}{4} (\cDB^{2} - 8{\bar R}) R {\bar M}_{\ad \bd} - \cDB_{( \ad} {\bar W}_{ \bd)}{}^{\gd \dd} {\bar M}_{\gd \dd} \non \\
&& + \frac{1}{6} \cDB_{( \ad} {\bar X}^{\gd} {\bar M}_{\bd) \gd} + \frac{1}{2} \cDB_{ ( \ad} \cD^{\g} G^{\d}{}_{\bd)} M_{\g \d} - \frac{1}{4} \cDB_{( \ad} {\bar X}_{\bd)} \mathbb{A} ~.
\eea
\end{subequations}
The torsion and curvature tensors are expressed in terms of the 
real vector $G_a$ and the complex superfields $R$, $X_\a$ and 
$W_{\a\b\g} = W_{(\a\b\g)}$, which have the $\sU(1)_{R}$ charges 
\bea
\label{U(1)Charges}
\mathbb{A} R = 2 R~, 
\qquad
\mathbb{A} X_{\a} = X_{\a}~,
\qquad 
\mathbb{A} W_{\a \b \g} = W_{\a \b \g}~.
\eea
and are covariantly chiral, 
\bea
 \cDB_{\ad} R = 0 ~, \qquad   \cDB_{\ad} X_{\a} = 0 ~,
 \qquad \cDB_{\ad} W_{\a \b \g} = 0 ~.
 \eea
These superfields obey the following Bianchi identities:
\begin{subequations}\label{Bianchi}
\bea
X_{\a} &=& \cD_{\a}R - \cDB^{\ad}G_{\a \ad} ~, \\
\cD^{\a} X_{\a} &=& \cDB_{\ad} {\bar X}^{\ad} ~,  \label{2.11b}\\
   \cD^{\g} W_{\a \b \g} &=& {\rm i} \cD_{(\a}{}^{\gd} G_{\b ) \gd} - \frac{1}{3} \cD_{(\a} X_{\b)} ~.
\eea
\end{subequations}
Equation \eqref{2.11b} means that $X_\a$ is the chiral field strength of an Abelian vector multiplet. 

In what follows we will use the notation $(\mathcal{M}^{4|4}, \cD)$
for the superspace $\mathcal{M}^{4|4}$ endowed with the geometry described.

%%%%%%%%%%%%%%%%%%%%%%%%%%%
%%%%%%%%%%%%%%%%%%%%%%%%%%%%

\subsection{Super-Weyl transformations}

In order for the above superspace
geometry  to describe conformal supergravity, the supergravity  
gauge group should include super-Weyl transformations, with 
the corresponding parameter $\Sigma$ being a real unconstrained scalar superfield. The defining property of these local rescalings is that they
preserve the structure of the algebra of covariant derivatives. 
In the infinitesimal case, the super-Weyl transformation is 
\begin{subequations}
\label{superWeylTf}
\bea
\delta_{\S}\cD_{\a} & = & \frac{1}{2} \S \cD_{\a} + 2 \cD^{\b} \Sigma M_{\b \a} + \frac{3}{2} \cD_{\a} 
\Sigma \mathbb{A} ~, \\
\d_{\S} \cDB_{\ad} & = & \frac{1}{2} \S \cDB_{\ad} + 2 \cDB^{\bd} \S {\bar M}_{\bd \ad} - 
\frac{3}{2} \cDB_{\ad} \S \mathbb{A} ~, \\
\d_{\S} \cD_{\a \ad} & = & \S \cD_{\a \ad} + {\rm i} \cD_{\a} \S \cDB_{\ad} 
+ {\rm i} \cDB_{\ad} \S \cD_{\a}  + {\rm i} \cDB_{\ad} \cD^{\b} \S  M_{\b \a} \non \\
&& + {\rm i} \cD_{\a} \cDB^{\bd} \S { \bar M}_{\bd \ad} - \frac{3}{4} {\rm i}  \left[ \cD_{\a} , \cDB_{\ad} \right]\S \mathbb{A} ~,
\eea
\end{subequations}
and the corresponding variations of the torsion and curvature superfields are
\begin{subequations}
\label{superWeylTfTorsions}
\bea
\d_{\S} R & = & \S R + \frac{1}{2} \cDB^{2} \S ~, \\
\d_{\S} G_{\a \ad} & = &  \S G_{\a \ad} + [ \cD_{\a} , \cDB_{\ad} ] \S ~, \\
\d_{\S} W_{\a \b \g} & = & \frac{3}{2} \S W_{\a \b \g} ~, \\
\d_{\S} X_{\a} & = & \frac{3}{2} \S X_{\a} - \frac{3}{2} (\cDB^{2} - 4 R) \cD_{\a} \S ~.
\eea
\end{subequations}
In Appendix \ref{Weyl_multiplet} we demonstrate that the
gauge transformations \eqref{gaugeTf} and \eqref{superWeylTf} allow us to choose 
a Wess-Zumino gauge in which the remaining fields  constitute the Weyl multiplet 
of conformal supergravity.

Consider a tensor superfield $U$ of  $\sU(1)_{R}$ charge $q_U$, 
\bea
{\mathbb A} U = q_U U~.
\eea
It is called primary if its super-Weyl transformation has the form
\bea
\d_\S U = \D_U \S U~,
\label{4.24}
\eea
for some parameter $\D_U$ called the dimension of $U$.
Given a primary superfield $ \J_{\a_1 \dots \a_n} = 
\J_{(\a_1 \dots \a_n)} $, which is covariantly chiral,  
$\bar \cD_\bd  \J_{\a_1 \dots \a_n} =0$, its dimensions and 
$\sU(1)_{R}$ charge are related to each other by 
\bea
 q_\J  = \frac 23 \D_\J ~.
\eea

For completeness, we also provide the finite super-Weyl transformation. It is 
\begin{subequations}\label{FinitesuperWeylTf}
\bea
\cD_{\a}' & = & \re^{\frac{1}{2} \S} \left( \cD_{\a} + 2 \cD^{\b} \Sigma M_{\b \a} + \frac{3}{2} \cD_{\a} \Sigma \mathbb{A} \right) ~, \\
\cDB_{\ad}' & = & \re^{\frac{1}{2} \S} \left( \cDB_{\ad} + 2 \cDB^{\bd} \S {\bar M}_{\bd \ad} - \frac{3}{2} \cDB_{\ad} \S \mathbb{A} \right) ~, \\
\cD_{\a \ad}' & = & \re^{\S} \Big( \cD_{\a \ad} + {\rm i} \cD_{\a} \S \cDB_{\ad} + {\rm i} \cDB_{\ad} \S \cD_{\a} + {\rm i} \left( \cDB_{\ad} \cD^{\b} \S  + 2 \cDB_{\ad} \S \cD^{\b} \S \right) M_{\b \a} \non \\ && + {\rm i} \left( \cD_{\a} \cDB^{\bd} \S + 2 \cD_{\a} \S \cDB^{\bd} \S \right) { \bar M}_{\bd \ad} \non \\
&&- 3{\rm i} \Big( \frac{1}{4} \left[ \cD_{\a} , \cDB_{\ad} \right] \S +  \cD_{\a} \S \cDB_{\ad} \S \Big) \mathbb{A} \Big) ~.~~~
\eea
\end{subequations}
The corresponding transformation laws for the torsion and curvature superfields are 
\begin{subequations}
\bea
R' & = & \re^{\S} \Big( R + \frac{1}{2} \cDB^{2} \S - ( \cDB \S )^{2} \Big) ~, \\
G_{\a \ad}' & = &  \re^{\S} \Big( G_{\a \ad} + [ \cD_{\a} , \cDB_{\ad} ] \S + 2 \cD_{\a} \S \cDB_{\ad} \S \Big) ~, \\
W_{\a \b \g}' & = & \re^{\frac{3}{2} \S} W_{\a \b \g} ~, \\
X_{\a}' & = & \re^{\frac{3}{2} \S} \Big( X_{\a} 
- \frac{3}{2} (\cDB^{2} - 4 R) \cD_{\a} \S \Big) ~.
\label{2.15d}
\eea
\end{subequations}
The super-Weyl tensor $W_{\a\b\g}$ and its conjugate $\bar W_{\ad\bd\gd}$ 
are the only torsion superfields which transform homogeneously under the 
super-Weyl group.

%%%%%%%%%%%%%%%%%%%
%%%%%%%%%%%%%%%%%%%%%%%%%%%

\subsection{From $\sU(1)$ superspace to the Grimm-Wess-Zumino geometry}

As pointed out above, the covariantly chiral spinor $X_\a$ is the field strength 
of an Abelian vector multiplet. It follows from \eqref{2.15d} that the super-Weyl gauge freedom allows us to choose the gauge 
\bea
X_\a =0~.
\label{2.16}
\eea
In this gauge the $\sU(1)_{R}$ curvature vanishes, in accordance with  
\eqref{algebra}, and therefore the $\sU(1)_{R}$ connection may be gauged away, 
\bea
\F_A =0~.
\eea
As a result, the algebra of covariant derivatives reduces \eqref{algebra} 
reduces to that describing the  GWZ geometry \cite{GWZ}.

Equation \eqref{2.15d}  tells us that imposing the condition $X_\a=0$ does not fix completely the super-Weyl freedom. The residual transformations are generated 
by parameters of the form 
\bea 
\S=\hf \big(\s +\bar \s \big) ~, \qquad \bar \cD_\ad \s =0~.
\label{2.18}
\eea
However, in order to preserve the  $\sU(1)_{R}$ gauge $\F_A=0$, 
every residual super-Weyl transformation \eqref{2.18} must be accompanied by a 
compensating $\sU(1)_{R}$ transformation with 
\bea
\r = \frac{3}{4} \ri \big(\bar \s - \s\big)~.
\eea
This leads to the transformation \cite{Siegel78,HT} 
\begin{subequations} 
\label{superweyl}
\bea
\d_\s \cD_\a &=& ( {\bar \s} - \hf \s)  \cD_\a + (\cD^\b \s) \, M_{\a \b}  ~, \\
\d_\s \bar \cD_\ad & = & (  \s -  \hf {\bar \s})
\bar \cD_\ad +  ( \bar \cD^\bd  {\bar \s} )  {\bar M}_{\ad \bd} ~,\\
\d_\s \cD_{\a\ad} &=& \hf( \s +\bar \s) \cD_{\a\ad} 
+\frac{\ri}{2} (\bar \cD_\ad \bar \s) \cD_\a + \frac{\ri}{2} ( \cD_\a  \s) \bar \cD_\ad \non \\
&& + (\cD^\b{}_\ad \s) M_{\a\b} + (\cD_\a{}^\bd \bar \s) \bar M_{\ad \bd}~.
\eea
\end{subequations}
The torsion tensors  transform
 as follows:
\begin{subequations} 
\bea
\d_\s R &=& 2\s R +\frac{1}{4} (\bar \cD^2 -4R ) \bar \s ~, \\
\d_\s G_{\a\ad} &=& \hf (\s +\bar \s) G_{\a\ad} +\ri \cD_{\a\ad} ( \s- \bar \s) ~, 
\label{s-WeylG}\\
\d_\s W_{\a\b\g} &=&\frac{3}{2} \s W_{\a\b\g}~.
\label{s-WeylW}
\eea
\end{subequations}

%%%%%%%%%%%%%%%%%%%%%%%%%%%%
%%%%%%%%%%%%%%%%%%%%%%%%%%%%

\section{Conformal isometries of curved superspace}\label{Section3}

Let $(\mathcal{M}^{4|4}, \cD) $ be a background superspace.
A real supervector field $\x= \x^B E_B$  is called conformal Killing if 
\bea
\label{master}
\left( \delta_{\mathcal{K}} + \delta_{\S} \right) \cD_{A} = 0 ~
\eea
for some Lorentz ($K^{\b \g}$), $R$-symmetry ($\r$) and super-Weyl ($\S$) parameters. 
Every solution to  \eqref{master}  defines a superconformal transformation 
of the superspace $(\mathcal{M}^{4|4}, \cD) $. 

%%%%%%%%%%%%%%%%%%%%%%%%%%%%%%%%%
%%%%%%%%%%%%%%%%%%%%%%%%%%%%%%%%%

\subsection{Implications of the superconformal Killing equation} 

Equation  \eqref{master}  contains nontrivial information. 
Choosing $A = \a$ in \eqref{master} and making use of 
the definition \eqref{gaugeTf} and \eqref{superWeylTf} in conjunction with the 
graded commutation relations \eqref{algebra}, 
we derive
\bea
\left(\delta_{\mathcal{K}} + \delta_{\S} \right) \cD_{\a} & = & \left( K_{\a}{}^{\b} - \cD_{\a} \xi^{\b} - \frac{\rm i}{2} \xi_{\a \bd} G^{\b \bd} - { \rm i} \delta_{\a}{}^{\b} \r  + \frac{1}{2}\delta_{\a}{}^{\b} \S \right) \cD_{\b} \non \\
&& + \left( \cD_{\a} {\bar \xi}^{\bd} + \frac{ \rm i}{2} \xi_{\a}{}^{\bd} \bar{R} \right) \cDB_{\bd} + 2 {\rm i} \left( {\bar \xi}^{\bd} \delta_{\a}{}^{\b} - \frac{\rm i}{4} \cD_{\a} \xi^{\b \bd} \right) \cD_{\b \bd} \non \\
&& - \left( \cD_{\a} K^{\b \g} + 4 \bar{R} \delta_{\a}{}^{( \b} \xi^{\g)} - \frac{\rm i}{2} \delta_{\a}{}^{(\b} \xi^{\g) \gd} \cDB_{\gd} \bar{R} - \frac{\rm i}{2} \xi_{\a \ad} \cD^{(\b} G^{\g) \ad} - 2 \delta_{\a}{}^{( \b} \cD^{\g)} \S \right) M_{\b \g} \non \\
&& - \left( \cD_{\a} {\bar K}^{\bd \gd} + {\rm i} \xi_{\a \ad} {\bar W}^{\ad \bd \gd} + \frac{\rm i}{6} \xi_{\a}{}^{( \bd} {\bar X}^{\gd)} \right) {\bar M}_{\bd \gd} \non \\
&& - {\rm i} \left( \cD_{\a} \r + \frac{1}{4} \xi_{\a \ad} {\bar X}^{\ad} + \frac{3 {\rm i}}{2} \cD_{\a} \S  \right) \mathbb{A} ~.
\eea 
Setting this to zero, we can read off the necessary conditions on our gauge and super-Weyl parameters for $\xi$ to be conformal Killing. 
These conditions can be split into two types. 
The first type provides expressions for the transformation parameters in terms of $\xi$
 \begin{subequations}
\label{type1constraints}
\bea
{\xi}^{\a} & = & - \frac{ \rm i}{8} \cDB_{\ad} \xi^{\a \ad} ~, \\
K_{\a \b} [ \xi ] & = & \cD_{(\a} \xi_{\b)} - \frac { \rm i}{2} \xi_{(\a}{}^{\ad} G_{\b) \ad} ~, \\
\r [ \xi ] & = & - \frac{\rm i}{4} \left(  \cD^\a \xi_\a - \cDB_\ad {\bar \xi} ^\ad\right) - \frac{1}{4} G^{\a \ad} \xi_{\a \ad} ~, \\
\label{sigmaXi}
\S [ \xi ] & = & - \frac{1}{2} \left( \cD^\a \xi_\a + \cDB_\ad {\bar \xi}^\ad \right) ~.
\eea
\end{subequations}
The second type
yields expressions for the spinor covariant derivatives of the parameters 
in terms of the original parameters and $\cD_\a \S [\x]$, including the following:
\begin{subequations}
\label{type2constraints1}
\bea
\cD_{\a} \xi_{\b} 
& = & \ve_{\a \b} \Big( \frac{\rm i}{4} G^{\g \gd} \xi_{\g \gd} + {\rm i} \r [ \xi ] - \frac{1}{2} \S [ \xi ] \Big) + K_{\a \b} [ \xi ] + \frac{\rm i}{2} \xi_{(\a}{}^{\ad} G_{\b)}{}_{\ad} ~, \\
\cD_{\a} {\bar \xi}_{\bd} & = & - \frac{\rm i}{2} \xi_{\a \bd} {\bar R} ~, \\
\cD_{\a} \xi_{\b \bd} & = & 4\ri \ve_{\a\b} \bar \x_\bd ~, \label{3.4c} \\
\cD_{\a} K^{\b \g} [ \xi ] & = & 2 \delta_{\a}{}^{(\b} \cD^{\g)} \S [ \xi ] - 4 \delta_{\a}{}^{(\b} \xi^{\g)} {\bar R} + \frac{\rm i}{2} \delta_{\a}{}^{(\b} \xi^{\g) \gd} \cDB_{\gd} {\bar R} \non \\
&& +\frac{\rm i}{2} \xi_{\a \ad} \cD^{(\b} G^{\g) \ad} ~, \\
\cD_{\a} {\bar K}^{\bd \gd} [ \xi ] & = & - {\rm i} \xi_{\a \ad} {\bar W}^{\ad \bd \gd} - \frac{ \rm i}{6} \xi_{\a}{}^{( \bd} {\bar X}^{\gd )} ~,\\
\cD_{\a} \r [ \xi] & = & - \frac{1}{4} \xi_{\a \ad} { \bar X}^{\ad} - \frac{3 \rm i}{2} \cD_{\a} \S [ \xi ] ~. 
\eea
\end{subequations}
The relations \eqref{type1constraints} tell us that all the
parameters are completely determined in terms of  $\xi^{a}$ and its covariant derivatives. As will be shown below, the relations \eqref{type2constraints1} 
imply that the the superalgebra of conformal Killing supervector fields is
finite dimensional. 

The above analysis shows that 
$\xi = \xi^{A} E_{A}$ is a conformal Killing supervector field if it has the form 
\begin{subequations} \label{newConformalKilling}
\bea
\xi^{A} &=& \left( \xi^{a} , - \frac{\rm i}{8} \cDB_{\bd} \xi^{\a \bd} , 
- \frac{\rm i}{8} \cD^{\b} \xi_{\b \ad} \right)~,
\eea
where $\x^a$ obeys the equation
\bea
 \cD_{(\a} \xi_{\b) \bd} = 0 \quad \Longleftrightarrow \quad \bar \cD_{(\ad} \x_{\b \bd)} =0~, 
 \label{newConformalKilling-b}
\eea
\end{subequations}
in accordance with \eqref{3.4c}.
Provided the equation \eqref{newConformalKilling-b}
 and definitions \eqref{type1constraints} hold, one may check that all the conditions 
\eqref{type2constraints1} are satisfied.
Equation \eqref{newConformalKilling-b} also implies that 
 $\x^a$ is covariantly linear, 
\bea
\label{xiLinear}
\left( \cD^{2} + 2 \bar{R} \right) \xi^a = 0 ~,
\label{3.66}
\eea
as well as the ordinary conformal Killing equation
\bea
\label{xiConformalKilling}
\cD_{(a} \xi_{b)} = \frac{1}{4} \eta_{a b} \cD^{c} \xi_{c} \quad \Longleftrightarrow
\quad \cD_{(\a (\ad } \x_{\b)\bd)} =0~.
\eea

Due to the relation $\{ \cD_{\a} , \cDB_{\ad} \} = -2{\rm i} \cD_{\a \ad} $, 
the equation \eqref{master} with $A = a$ is automatically satisfied 
once \eqref{master} with $A = \a$ holds. 
Still the implications of the equation \eqref{master} with $A = a$
prove to be very useful for computations, and we spell them out here:
\begin{subequations}
\label{type2constraints2}
\bea
\cD_{\a \ad} \xi^{\b} &=& - {\rm i} \xi_{\a} G^{\b}{}_{\ad} - {\rm i} \delta_{\a}{}^{\b} {\bar \xi}_{\ad} R - \frac{1}{4} \xi_{\a}{}^{\bd} \cDB_{(\ad} G^{\b}{}_{\bd)} + \frac{1}{4} \delta_{(\a}{}^{\b} \xi^{\g}{}_{\ad} \cD_{\g)} R \non \\ 
&& + \frac{1}{2} \xi^{\g}{}_{\ad} W_{\a \g}{}^{\b} + \frac{1}{12} \d_{(\a}{}^{\b} X_{\g)} \xi^{\g}{}_{\ad} + {\rm i} \delta_{\a}{}^{\b} \cDB_{\ad} \S [ \xi ] ~, \label{CKSpinor} \\
\cD_{\a \ad} \xi^{\b \bd} & = & - {\rm i} \d_{( \ad}{}^{\bd} \xi_{\a}{}^{\gd} G^{\b}{}_{\gd )} + {\rm i} \d_{( \a}{}^{\b} \xi^{\g}{}_{\ad} G_{\g )}{}^{\bd} - 2 \d_{\ad}{}^{\bd} K_{\a}{}^{\b} [ \xi ] - 2  \d_{\a}{}^{\b} {\bar K}_{\ad}{}^{\bd} [ \xi ] \label{CKVector} \non \\
&& - 2 \d_{\a}{}^{\b} \d_{\ad}{}^{\bd} \S [\xi], \\
\cD_{\a \ad} K^{\b \g} [ \xi ] & = & {\rm i} \xi_{\a} \cD^{(\b} G^{\g)}{}_{\ad} + {\rm i} \d_{\a}{}^{(\b} \xi^{\g)} \cDB_{\ad} {\bar R} + \frac{\rm i}{3} \d_{\a}{}^{( \b} {\bar \xi}_{\ad} X^{\g ) } - 2 {\rm i} {\bar \xi}_{\ad} W_{\a}{}^{\b \g} \non \\
&& + \frac{1}{4} \xi_{\a}{}^{\bd} \cDB_{(\ad} \cD^{(\b} G^{\g)}{}_{\bd)} + \frac{1}{8}  \d_{\a}{}^{(\b} \xi^{\g)}{}_{\ad} ( \cD^{2} - 8 \bar{R} ) R + \frac{1}{2} \xi^{\l}{}_{\ad} \cD_{(\a} W_{\l)}{}^{\b \g} \non \\
&& - \frac{1}{12} \xi^{\l}{}_{\ad} \d_{(\a}{}^{(\b} \cD_{\l)} X^{\g)} + {\rm i} \d_{\a}{}^{( \b} \cDB_{\ad} \cD^{\g)} \S [ \xi ] ~, \\
\cD_{\a \ad} \r [ \xi ]& = & \frac{1}{2} {\bar \xi}_{\ad} X_{\a} - \frac{1}{2} \xi_{\a} { \bar X}_{\ad} + \frac{\rm i}{8} \xi_{\a}{}^{\bd} \cDB_{(\ad} {\bar X}_{\bd)} + \frac{\rm i}{8} \xi^{\b}{}_{\ad} \cD_{(\a} X_{\b)} \non \\
&& - \frac{3}{4} [ \cD_{\a} , \cDB_{\ad} ] \S [ \xi ] ~.
\eea
\end{subequations}
We emphasise once more that these identities may be derived by making use of \eqref{type2constraints1}.

%%%%%%%%%%%%%%%%%%%%%%%%%%%%%%
%%%%%%%%%%%%%%%%%%%%%%%%%%%%%%
%

\subsection{The superconformal algebra}

It follows from  \eqref{master} that commuting two superconformal transformations 
of $(\mathcal{M}^{4|4}, \cD) $ 
results in another transformation of the same type, 
\begin{subequations}
\bea
\left[ \d_{\mathcal{K}[\xi_{2}]} + \d_{\S[\x_{2}]} , \d_{\mathcal{K}[\xi_{1}]} 
+ \d_{\S[\x_{1}]} \right] \cD_{A} &=& 
\left( \d_{\mathcal{K}[\xi_3]} + \d_{\S[\x_3]} \right) \cD_{A} = 0 ~,\\
 \mathcal{K}[\xi_3] &:=& \Big[ \mathcal{K}[\xi_{2}] , \mathcal{K}[\xi_{1}] \Big] ~.
\eea
\end{subequations}
This means that  the set of all conformal Killing supervector fields forms a Lie superalgebra, the superconformal algebra of $(\mathcal{M}^{4|4}, \cD) $. 

It is of interest to derive the explicit expressions for $\S[\x_3]$ and $\x^a_3$ in terms 
of $\x_1^a$ and $\x_2^a$. 
A routine calculation gives
\bea
\left[ \mathcal{K}[\xi_3] , \cD_{A} \right] 
+ \left[ \mathcal{K}[\xi_{2}] , \d_{\S[\x_{1}]}\cD_{A} \right] - \left[ \mathcal{K}[\xi_{1}] , \d_{\S[\x_{2}]} \cD_{A} \right] = 0 ~.
\eea
Specialising here to the $A = \a$ case and extracting the super-Weyl parameter, we find
\bea
\S[\x_3]= \xi_{2}^{A} \cD_{A} \S[\x_{1}] - \xi_{1}^{A} \cD_{A} \S[\x_{2}] ~.
\eea
For the vector component $\x^a_3 $ we obtain
\bea
\x^{\a\ad}_3 &=& -\hf  \x^{\b\bd}_1 \cD_{\b \bd} \x^{\a\ad}_2 
- \frac{\ri}{16}  \bar \cD_\bd \x^{\a\bd}_1 \cD_\b \x^{\b \ad}_2 
+\frac{\ri}{2}  \x_{1}^{\a\bd} \x_2^{\b \ad} G_{\b \bd}  ~-~ \big(1 \leftrightarrow 2\big)
~.
\eea
One may check that $\x^{\a\ad}_3$ obeys 
the superconformal Killing equation \eqref{newConformalKilling-b}.

The superconformal algebra of $(\mathcal{M}^{4|4}, \cD) $ turns out to be finite dimensional, 
and its dimension does not exceed that of the $\cN=1$ superconformal group $\sSU(2,2|1)$. 
In order to prove this claim, 
we introduce the following set of parameters: 
\bea
\label{conformalparameters}
{\bm \X} := \Big\{ \xi^{A} , \,K^{\a \b} [ \xi ] , \,\bar{ K }^{\ad \bd} [ \xi ] , \,\r [ \xi ] , \, \S [ \xi ] , \,\cD_{A} \S [ \xi ] \Big\}. 
\eea
It is not difficult to demonstrate that $\cD_A \bm \X $ is 
a linear combination of the elements of 
\eqref{conformalparameters}.
Actually, it suffices to show that $\cD_{\a} \bm \X$ satisfies this property, as the general case immediately follows.
Due to the relations \eqref{type2constraints1} and \eqref{type2constraints2}, 
we only need to analyse $\cD_\a \cD_B \S[\x]$.
Direct calculations give
\begin{subequations} \label{3.14}
\bea
\cD_{\a} \cD_{\b} \S  [ \xi ]& = &  \frac{1}{2} \ve_{\a \b} \cD^{2} \S 
=  -\ve_{\a \b} \Big( \big(  \S [ \xi ] -2 {\rm i} \r [ \xi ]  \big) \bar{R} +  \xi^{c} \cD_c R + {\bar \xi} \cDB {\bar R} \Big) ~, \\
\cD_{\a} \cDB_{\bd} \S [ \xi ] & = &  - {\rm i} \cD_{\a \bd} \S [ \xi ] 
+ \frac{1}{2} {\bar \xi}^{\gd} \cDB_{\gd} G_{\a \bd} 
- \frac{1}{2} \xi^{\g} \cD_{\g} G_{\a \bd}  \non \\
&& - \frac{1}{4} \xi_{\a}{}^{\gd} \cD^{\g}{}_{(\bd} G_{\g \gd)}
+ \frac{1}{4} \xi^{\g}{}_{\bd} \cD_{(\a}{}^{\gd} G_{\g) \gd} 
+ \frac{1}{4} \xi^{\g \gd} \cD_{\a \bd} G_{\g \gd} \non \\
&& - \frac{1}{2} K_{\a}{}^{\g} [ \xi ] G_{\g \bd} 
- \frac{1}{2} {\bar K}_{\bd}{}^{\gd} [ \xi ] G_{\a \gd} 
- \frac{1}{2} G_{\a \bd} \S [ \xi ] ~, \\
%%%%%%%%%%%%%%%%%%%%%%%%%%%%
\cD_{\a} \cD_{\b \bd} \S [ \xi ] & = & \ve_{\a \b} \bigg[ \Big( \frac{\rm i}{8} \d^{\ad}{}_{\bd} \S [ \xi ] + \frac{1}{12} \d^{\ad}{}_{\bd} \r [ \xi ] + \frac{\rm i}{12} {\bar K}^{\ad}{}_{\bd} [ \xi ] \Big) \left( {\bar X}_{\ad} + 3 {\bar \cD}_{\ad} {\bar R} \right) + {\rm i} {\bar R} \cDB_{\bd} \S [ \xi ] \non \\ 
&& - \frac{\rm i}{2} G^{\g}{}_{\bd} \cD_{\g} \S [ \xi ] - \frac{\rm i}{12} {\bar \xi}^{\ad} \cDB_{ ( \ad} {\bar X}_{\bd )} + \frac{7 \rm i}{24} {\bar \xi}_{\bd} \cD X - \frac{ \rm i}{8} {\bar \xi}_{\bd} \cDB^{2} {\bar R} + \frac{ 3 \rm i}{4} \xi^{\g} G_{\g \bd} {\bar R} \non \\
&& - \frac{\rm i}{8} \xi^{\g} \cD^{2} G_{\g \bd} + \frac{1}{12} \xi^{\g \ad} \cD_{\g} \cDB_{ ( \ad} {\bar X}_{\bd)} - \frac{1}{48} \xi^{\g}{}_{\bd} \cD^{2} X_{\g} - \frac{\rm i}{8} \xi^{\g \ad} \cD_{\g \ad} ( \cDB_{\bd} {\bar R} - {\bar X}_{\bd} ) \bigg]
\non \\
&&  + \bigg[ 6 \Big( \frac{\rm i}{8} \d^{\ad}{}_{\bd} \S [ \xi ] + \frac{1}{12} \d^{\ad}{}_{\bd} \r [ \xi ] + \frac{\rm i}{12} {\bar K}^{\ad}{}_{\bd} [ \xi ] \Big) \cD_{( \a} G_{\b) \ad} + {\rm i} K_{(\a}{}^{\g} [ \xi ] \cD_{\b)} G_{\g \bd} \non \\
&& - \frac{\rm i}{2} K_{\a \b} [ \xi ] \left( \cDB_{\bd} \bar{R} - {\bar X}_{\bd} \right) - \frac{\rm i}{2} {\bar \xi}^{\ad} \cDB_{\ad} \cD_{(\a} G_{\b) \bd} - \frac{\rm i}{4} \xi_{(\a} \left( \cD^{2} + 2 \bar{R} \right) G_{\b) \bd} \non \\
&& - \frac{1}{12} \xi_{(\a (\ad} G_{\b) \bd)} {\bar X}^{\ad} - \frac{1}{12} \xi_{(\a \ad} G_{\b)}{}^{\ad} {\bar X}_{\bd} - \frac{\rm i}{4} \xi^{\g \ad} \cD_{\g \ad} \cD_{( \a} G_{\b) \bd}
\bigg]
\eea
\end{subequations}
Thus, we have demonstrated that the superconformal algebra is  finite dimensional.

%%%%%%%%%%%%%%%%%%%%%%%%%%
%%%%%%%%%%%%%%%%%%%%%%%%%%

\subsection{Conformally related superspaces}\label{section3.3}

Let $(\mathcal{M}^{4|4}, \cD) $ and 
$(\mathcal{M}^{4|4}, \hat{\cD}) $ be two supergravity backgrounds. 
We say that the two superspaces are conformally related 
if their covariant derivatives  $\hat{\cD}_{A}$
and $\cD_{A}$ are
related to each other  by a finite super-Weyl transformation \eqref{FinitesuperWeylTf},
\begin{subequations} \label{ConRel}
\bea
\hat{\cD}_{\a} & = & \re^{\frac{1}{2} \S} \left( \cD_{\a} + 2 \cD^{\b} \Sigma M_{\b \a} + \frac{3}{2} \cD_{\a} \Sigma \mathbb{A} \right) ~, \\
\hat{ \bar{\cD}}_\ad & = & \re^{\frac{1}{2} \S} \left( \cDB_{\ad} + 2 \cDB^{\bd} \S {\bar M}_{\bd \ad} - \frac{3}{2} \cDB_{\ad} \S \mathbb{A} \right) ~, \\
{\hat \cD}_{\a\ad} & = & \frac{\ri}{2} \big\{ {\hat \cD}_\a, \hat{\bar {\cD}}_\ad \big\} ~.
\eea
\end{subequations}
 These  superspaces  prove to have the same conformal Killing supervector fields, 
\bea
\xi = \xi^{A} E_{A} = \hat{\xi}^{A} \hat{E}_{A} ~,
\eea
where the components $\hat{\xi}^{A} $ are given by
\bea
\hat{\xi}^{\a \ad} = {\rm e}^{- \S} \xi^{\a \ad} , \qquad 
\hat{\xi}^{\a} = {\rm e}^{- \frac{1}{2} \S} \left( \xi^{\a} + \frac{\rm i}{2} \xi^{\a \bd} \cDB_{\bd} \S \right) ~.
\label{3.19}
\eea
The transformed supervector field $\hat{\xi}^{A}$ also satisfies \eqref{newConformalKilling} (in the new basis), thus it is a conformal Killing vector
\bea
\hat{\xi}^{A} = \Big( \hat{\xi}^{a} , - \frac{\rm i}{8} \hat{\bar \cD}_{\bd} \hat{\xi}^{\a \bd} , 
- \frac{\rm i}{8} \hat{\cD}^{\b} {\xi}_{\b \ad} \Big)~, 
\qquad \hat{\cD}_{(\a} \hat{\xi}_{\b) \bd} = 0 ~.
\eea
One can relate the remaining parameters generating conformal isometries in each geometry in a simple way
\begin{subequations}
\bea
\S [ \hat{ \xi } ] & = & \S [ \xi ] - \xi^{A} \cD_{A} \S [ \xi ] ~, \\
K_{\a \b} [ \hat{\xi} ] & = & K_{\a \b} [ \xi ] + 2 \cD_{( \a} \S \xi_{\b )} + \frac{\ri}{2} \cDB_{\ad} \cD_{( \a} \S \xi_{\b)}{}^{\ad} ~, \\
\r [ \hat{ \xi } ] & = & \r [ \xi ] + \frac{3 \ri}{2} \cD^{\a} \xi_{\a} - \frac{3 \ri}{2} \cDB_{\ad} \S \bar{\xi}^{\ad}
- \frac{3}{8} [ \cD^{\a} , \cDB^{\ad} ] \S \xi_{\a \ad} ~.
\eea
\end{subequations}
It then follows that the gauge transformation is identical in these two geometries $\mathcal{K}[\xi] = \mathcal{K}[\hat{\xi}]$; it is a super-Weyl invariant operator.

%%%%%%%%%%%%%%%%%%%%%%%%%%%%%%%%
%%%%%%%%%%%%%%%%%%%%%%%%%%%%%%%%%

\subsection{Superconformal field theory}

Let $\vf^i$ be the dynamical superfield variables describing a matter system coupled to conformal supergravity. The matter action is required to be invariant under the super-Weyl 
transformations \eqref{superWeylTf} accompanied by certain transformations of the matter superfields of the form 
\bea
\d_\S \vf^i =\D_{(i)} \S \vf^i~, 
\eea
where $\D_{(i)}$ denotes the dimension of $\vf^i$. 
In general, the matter action includes two terms 
\bea
S&=&\int \rd^4x\rd^2\q\rd^2\qb  \,E \,\cL 
+ \left\{ \int\rd^4x\rd^2\q\, \cE \,\cL_{\rm c} +{\rm c.c.} \right\}
~, \qquad E^{-1} = {\rm Ber}(E_A{}^M)~,
\label{SCFT}
\eea
with $\cE$ being the so-called chiral density. 
Here the full superspace Lagrangian $\cL$ is a primary real scalar superfield of dimension $+2$, while $\cL_{\rm c} $ is a primary covariantly chiral superfield,  
$\bar \cD_\ad \cL_{\rm c} =0$,
of dimension $+3$, 
\bea
\d_\S \cL =2\S \cL~, \qquad
 \d_\S \cL_{\rm c} &=&3\S \cL_{\rm c} ~.
\eea
It should be pointed out that the full superspace measure $E$ and the 
chiral density $\cE$ have the following super-Weyl transformation laws
\bea
\d_\S E = - 2 \S E~, \qquad \d_\S \cE = - 3 \S \cE ~.
\eea

The chiral density can be naturally defined using the prepotential solution of the supergravity 
constraints given in \cite{GGRS}. 
It can also be obtained using the general formalism of integrating out fermionic dimensions, which was developed in \cite{KT-M-2008-2}.
Probably the simplest definition of the chiral action 
\bea
S_{\rm c} = \int\rd^4x\rd^2\q\, \cE \,\cL_{\rm c} ~,
\label{chiralAc}
\eea
is described in Appendix \ref{AppendixA}. 
The full superspace action   can be represented 
as an integral over the chiral subspace, 
\bea
\int\rd^4x\rd^2\q\rd^2\qb\, E \,\cL
= -\frac 14 \int\rd^4x\rd^2\q\, \cE \,\big(\cDB^2-4R\big)
\cL~.
\label{full-chiral}
\eea

In the case of a fixed supergravity background, the matter action \eqref{SCFT} 
is invariant under superconformal transformations of the form 
\bea
\d_\x \vf^i =  \mathcal{K}[\xi] \vf^i +\D_{(i)} \S[\x] \vf^i~,
\label{3.28}
\eea
where  $\x^A$ is an arbitrary conformal Killing supervector field 
of the background curved superspace $(\mathcal{M}^{4|4}, \cD) $. 

An important example of a superconformal field theory in curved superspace 
is the massless Wess-Zumino model 
\bea
S[\f, \bar \f] &=&\int \rd^4x\rd^2\q\rd^2\qb  \,E \,\bar \f \f 
+ \left\{ \frac{\l}{3!} \int\rd^4x\rd^2\q\, \cE \,\f^3 +{\rm c.c.} \right\}~,
\qquad \bar \cD_\ad \f =0~, ~~
\label{3.29}
\eea
with $\l $ a coupling constant. 
Here the chiral scalar $\f$ is primary and of dimension $+1$.

%%%%%%%%%%%%%%%%%%%%%%%%%%%%%%%%
%%%%%%%%%%%%%%%%%%%%%%%%%%%%%%%%

\subsection{Superconformal sigma models}

A nontrivial example of a superconformal field theory on 
 $(\mathcal{M}^{4|4}, \cD) $ is a nonlinear sigma model. 
 The target spaces of superconformal sigma models are K\"ahler cones \cite{GR}.
Let us recall what this means.
Consider a K\"ahler manifold $(\cN, g_{\m\n}, J^\m{}_\n )$, where $\m,\n=1,\dots, 2n$,  
and introduce local complex coordinates
$\f^i$ and their conjugates $\bar \f^{\bar i}$, in which the complex structure 
$J^\m{}_\n$ is diagonal. 
It is called a K\"ahler cone \cite{GR} if it possesses
 a homothetic conformal Killing vector
\bea
\c = \c^i \frac{\pa}{\pa \f^i} + {\bar \c}^{\bar i}  \frac{\pa}{\pa {\bar \f}^{\bar i}}
\equiv \c^\m \frac{\pa}{\pa \vf^\m} ~,
\eea
with the following properties:
\bea
\nabla_\n \c^\m = \d_\n{}^\m \quad \Longleftrightarrow \quad 
\nabla_j \c^i = \d_j{}^i~, \qquad 
\nabla_{\bar j} \c^i = \pa_{\bar j} \c^i = 0~,
\label{hcKv}
\eea
which show, in particular, that  $\c $ is holomorphic. 
In terms of the scalar field $K:= { g}_{i \bar j} \, \c^i {\bar \c}^{\bar j} $ on the target space,
these properties imply that 
\bea
\qquad \c_i = {g}_{i \bar j} \,{\bar \c}^{\bar j} = \pa_i { K}~,\qquad 
{g}_{i \bar j} = \pa_i \pa_{\bar j} { K}~,
\label{hcKv-pot}
\eea
and therefore
\bea
\c^i(\f) \pa_i K (\f, \bar \f)  = K (\f , \bar \f) ~.
\eea
The real function $K (\f , \bar \f) $ is a globally defined K\"ahler potential.
Associated with $\c$ is the $\sU(1)$ Killing vector field 
\bea 
V^\m = J^\m{}_\n \c^\n~, \qquad \nabla_\m V_\n + \nabla_\n V_\m =0~.
\eea
Local complex coordinates $\f^i$ can always be
chosen such that $\c^i(\f) = \f^i$.

Consider the following nonlinear $\sigma$-model 
\bea
S&=&\int \rd^4x\rd^2\q\rd^2\qb\, E \, K\big(\f , {\bar \f} \big)~, 
\qquad {\bar \cD}_{\dot \a} \f^i =0~,
\label{N=1sigma-model}
\eea
where the action of the $\sU(1)_R$ generator on $\f^i$ is defined as
\bea
 {\mathbb A} \f^i =\frac 23 \c^i (\f) ~.
 \eea 
 The action is invariant under super-Weyl transformations
 \bea
 \d_\S \f^i = \S  \c^i (\f) ~.
 \eea
In the case of a fixed supergravity background, the matter action \eqref{SCFT} 
is invariant under superconformal transformations of the form 
\bea
\d_\x \f^i =  \mathcal{K}[\xi] \f^i + \S[\x] \c^i(\f)~,
\eea
where  $\x^A$ is an arbitrary conformal Killing supervector field 
of the background curved superspace $(\mathcal{M}^{4|4}, \cD) $. 

%%%%%%%%%%%%%%%%%%%%%%%%%%%%%%%%%%%%%
%%%%%%%%%%%%%%%%%%%%%%%%%%%%%%%%%%%%%%%

\section{Conformal Killing tensor superfields}\label{section4}

As discussed in Section \ref{Section3}, every conformal Killing supervector field 
$\x^A$ of the background curved superspace $(\mathcal{M}^{4|4}, \cD) $
 is determined by its vector component $\x^a$, which is real and constrained by 
\bea
 \cD_{(\a} \xi_{\b) \bd} = 0   \quad \Longleftrightarrow \quad \bar \cD_{(\ad} \xi_{\b \bd)} = 0 ~.
 \label{4.1}
 \eea
 It follows from \eqref{3.19} that $\x_{\a\ad} $ has  the super-Weyl transformation law \bea
 \d_\S \x_{\a\ad} = - \S \x_{\a\ad}~,
 \eea
 which  is uniquely determined by requiring 
  equations \eqref{4.1} to be super-Weyl invariant. 
 This construction admits nontrivial generalisations. 
 
 %%%%%%%%%%%%%%%%%%%%%%%%%%%%%%%%%%%%
 %%%%%%%%%%%%%%%%%%%%%%%%%%%%%%%%%%%%
 
\subsection{Definitions} 
 
Let $m $ and $ n$ be non-negative integers.
A primary tensor superfield $\ell_{\a(m) \ad(n)}$ on $(\mathcal{M}^{4|4}, \cD) $
is called conformal Killing if it obeys the constraints\footnote{These constraints 
can be naturally lifted to the conformal superspace of \cite{Butter4DN=1}.} 
\begin{subequations}\label{4.3}
\bea
\cD_{(\a_1} \ell_{\a_2 \dots \a_{m+1} ) \ad(n)} &=& 0 \quad \implies \quad 
\big(\cD^ 2 + 2 m \bar{R} ) \ell_{\a(m) \ad(n)}=0~,
\label{4.3a} \\
\bar \cD_{(\ad_1} \ell_{\a(m) \ad_2 \dots \ad_{n+1} ) } &=& 0 \quad \implies \quad
\big( \cDB^{2} + 2 n R ) \ell_{\a(m) \ad(n)} = 0 ~.
\label{4.3b}
\eea
\end{subequations}
These conditions imply the following transformation properties:
\begin{subequations}
\bea
\d_\S \ell_{\a(m) \ad(n)} &=& - \hf (m+n) \S \,\ell_{\a(m) \ad(n)}~, \\
{\mathbb A} \ell_{\a(m) \ad(n)} &=&  -\frac 13 (m -n) \ell_{\a(m) \ad(n)}~.
\eea
\end{subequations}
If $m=n$, then $ \ell_{\a(n) \ad(n)} $ is neutral with respect to 
the $R$-symmetry group $\sU(1)_{R}$, and therefore it is consistent to 
restrict  $ \ell_{\a(n) \ad(n)} $ to be real. Another special choice is $n=0$, 
in which case $\ell_{\a(m)} $ is covariantly chiral, 
$\bar \cD_{\ad} \ell_{\a(m) } = 0$.

The constraints \eqref{4.3} provide a natural generalisation of the concept  of 
a {\it conformal} Killing tensor field $L_{\a(m) \ad (n)}$ on a curved spacetime $\cM^4$ \cite{Penrose}.\footnote{Penrose and Rindler \cite{Penrose} called  $L_{\a(m) \ad (n)}$
a Killing spinor.}
By definition,  $L_{\a(m) \ad (n)}$  is a primary field which obeys the equation
\bea
\nabla_{(\a_1}{}^{(\ad_1} L_{\a_2 \dots \a_{m+1}) }{}^{\ad_2 \dots \ad_{n+1})} =0~,
\label{4.55}
\eea 
where $\nabla_{\a\ad}$ is the torsion-free Lorentz-covariant derivative. The condition 
that $L_{\a(m) \ad (n)}$  is  primary means that it changes homogeneously under a Weyl 
transformation 
\bea
\d_\s \nabla_a = \s \nabla_a - \nabla^b\s M_{ab}~,
\eea
with $\s(x) $ the Weyl parameter. The unique Weyl transformation law of
 $L_{\a(m) \ad (n)}$, which is compatible with the constraint \eqref{4.55},  is
 \bea
  \d_\s L_{\a(m) \ad (n)} = -\hf (m+n) \s  L_{\a(m) \ad (n)}~.
  \eea

Given two conformal Killing tensor superfields 
$\ell_{\a(m) \ad(n)}$ and $\ell_{\a(p) \ad(q)}$
on $(\mathcal{M}^{4|4}, \cD) $, 
their symmetric  product 
\bea
\ell_{\a(m+p) \ad(n+q)}:=  \ell_{(\a_1 \dots \a_m ( \ad_1 \dots \ad_n}
\ell_{\a_{m+1} \dots \a_{m+p} ) \ad_{n+1} \dots \ad_{n+q})} ~,
 \eea
 is also conformal Killing. This operation allows one to generate new conformal Killing tensor superfields from given ones. 

Constraints \eqref{4.3} naturally occur in the framework of 
conformal higher-spin gauge supermultiplets
\cite{KMT,KP2}. For $m\geq n >0$ such a supermultiplet
is described by an unconstrained primary prepotential
$\U_{\a (m) \ad (n)} $ defined modulo gauge transformations 
\bea
 \d_{ \L, \z} \U_{\a (m) \ad (n)} 
 =  
 \cD_{(\a_1}\z_{\a_2 \dots \a_m)\ad_1 \dots \ad_{n}} 
+\bar \cD_{(\ad_1} \L_{\a_1 \dots \a_m \ad_2 \dots \ad_{n} )}
 \ ,
\label{4.5}
\eea
with unconstrained  primary gauge parameters 
$\z_{\a(m-1)\ad(n)}$ and $ \L_{\a (m) \ad (n-1)} $. 
In the  $m > n=0$ case, the conformal gauge supermultiplet 
is described by an unconstrained primary prepotential
$\U_{\a (m) } $ defined modulo gauge transformations 
\bea
 \d_{ \z, \l} \U_{\a (m) } 
 = \cD_{(\a_1}\z_{\a_2 \dots \a_m)}  + \l_{\a(m)} ~, 
 \qquad \bar \cD_\bd \l_{\a(m)} =0~.
\label{4.6}
\eea
Now, if we look for special gauge parameters 
$\z_{\a(m-1)\ad(n)}$ and $ \L_{\a (m) \ad (n-1)} $
such that the variation \eqref{4.5} vanishes, 
$ \d_{ \L, \z} \U_{\a (m) \ad (n)} =0 $, then 
$\ell_{\a(m)\ad(n)} :=  \cD_{(\a_1}\z_{\a_2 \dots \a_m)\ad_1 \dots \ad_{n}} $
is a solution to the constraints  \eqref{4.3}.

A higher-spin interpretation exists also for the conformal Killing tensors \eqref{4.55}. 
We recall that  a conformal
higher-spin gauge field $h_{\a(m+1)\ad(n+1)}$ is 
a primary field defined modulo gauge transformations \cite{FT}
\bea
\delta_{\lambda}h_{\a(m+1)\ad(n+1)}
=\nabla_{(\a_1(\ad_1}\lambda_{\a_2\dots\a_{m+1})\ad_2\dots\ad_{n+1})}~,
\eea 
 where the gauge parameter $\lambda_{\a(m)\ad(n)}$ is also primary. 
The conformal  Killing tensors \eqref{4.55} correspond to those values of the gauge parameter $\lambda_{\a(m)\ad(n)}$ 
which leave the gauge field invariant, $\delta_{L}h_{\a(m+1)\ad(n+1)} =0$.

The importance of the conformal Killing superfields $\ell_{\a(m) \ad(n)}$ introduced
is that they generate symmetries of dynamical systems on $(\mathcal{M}^{4|4}, \cD) $.
We have seen that the $\cN=1$ superconformal transformations 
are described by $\ell_{\a\ad}$. In the next subsection, we introduce
various important conformal supercurrents and describe their interplay
with conformal Killing tensor superfields. Following this, in sections
\ref{section4.3} and \ref{section4.4} we show that extended superconformal transformations 
are formulated in terms of $\ell_\a$ and its conjugate.
 Then in section \ref{section4.5}, it  will be demonstrated that higher-rank analogues of $\ell_{\a\ad}$,
 the conformal Killing tensor superfields $\ell_{\a (n) \ad (n) }$, generate symmetries 
 of the massless Wess-Zumino operator.

%%%%%%%%%%%%%%%%%%%%%%%%%%%%%%%%
%%%%%%%%%%%%%%%%%%%%%%%%%%%%%%%%

\subsection{Conserved current supermultiplets} \label{subsection4.22}

When considering conformal field theories on ${\mathbb R}^{d-1,1}$,
  a well-known procedure 
exists to generate  conserved conformal currents by making use of 
 a symmetric, traceless and conserved  energy-momentum tensor 
$T^{ab}$
\bea
T^{ab}= T^{ba}~, \qquad \eta_{ab}T^{ab}=0~, \qquad \pa_b T^{ab}=0~,
\eea
with $\eta_{ab} $ the Minkowski metric. 
Given   a conformal Killing vector field $\x= \x^a \pa_a$,
\bea
\pa_a \x_b + \pa_b \x_a = \frac{2}{d}
\eta_{ab} \pa_c \x^c
~,
\eea
the following vector field 
\bea
\label{cc}
j^a[\x]= T^{ab}\x_b
\eea
is conserved, $  \pa_a j^a =0$. The construction is naturally generalised to a curved space. 
It also has a higher-spin extension  \cite{Mikhailov}.
Here we will present  supersymmetric extensions of these constructions building, in part, on the earlier 
work \cite{KKT}.

Let $m $ and $ n$ be positive integers.
A primary tensor superfield $J^{\a(m) \ad(n)}$ on $(\mathcal{M}^{4|4}, \cD) $
is called a conformal supercurrent of valence $(m,n)$ 
 if it obeys the constraints
\begin{subequations}\label{supercurrent} 
\bea
\cD_{\b} J^{\b\a (m-1)\ad(n)} &=& 0 \quad \implies \quad \big( \cD^{2} - 2 ( m + 2 ) \bar{R} \big) J^{\a(m) \ad(n)} = 0 ~, 
\label{supercurrent-a} 
\\
\bar \cD_{\bd} J^{\a(m) \bd \ad(n-1) } &=& 0 \quad \implies \quad \big( \cDB^{2} - 2 ( n + 2 ) R \big) J^{\a(m) \ad(n)} = 0 ~.
\label{supercurrent-b} 
\eea
\end{subequations}
These conditions imply the following superconformal transformation properties:
\begin{subequations}\label{s-transformation}
\bea
\d_\S J^{\a(m) \ad(n)} &=& \Big(2+ \hf (m+n) \Big)\S \,J^{\a(m) \ad(n)}~, \\
{\mathbb A} J^{\a(m) \ad(n)} &=&  \frac 13 (m -n) J^{\a(m) \ad(n)}~.
\eea
\end{subequations}
If $m=n$, then $ J^{\a(n) \ad(n)} $ is neutral with respect to 
the $R$-symmetry group $\sU(1)_{R}$, and therefore it is consistent to 
restrict  $ J^{\a(n) \ad(n)} $ to be real. The $m=n=1$ case corresponds to the ordinary 
conformal supercurrent \cite{FZ}. The case $m=n>1$ was first described in Minkowski superspace in \cite{HST81} and extended to AdS superspace in \cite{BHK}.

In the case $m >n =0$, the constraints \eqref{supercurrent} should be replaced with 
\begin{subequations}\label{supercurrent2} 
\bea
\cD_{\b} J^{\b\a (m-1)} &=& 0 \quad \implies \quad \big( \cD^{2} - 2 ( m + 2 ) \bar{R} \big) J^{\a(m)} = 0 ~, 
\label{supercurrent2-a} 
\\
(\bar \cD^2 -4R)  J^{\a(m)  } &=& 0~.
\label{supercurrent2-b} 
\eea
\end{subequations}
The superconformal transformation properties of $J^{\a(m)}$ are obtained from 
\eqref{s-transformation} by setting $n=0$. 
The case $n=1$ was first considered in \cite{KT}, where it was shown that the spinor supercurrent $J^\a$ naturally originates from the reduction of the conformal $\cN=2$ supercurrent \cite{Sohnius79}
to $\cN=1$ superspace. 

Finally, for $m=0$  
the constraints \eqref{supercurrent2} should be replaced with 
\begin{subequations}
\bea\label{supercurrent3} 
(\cD^2 - 4\bar R) J &=&0~, \\
(\bar \cD^2 -4R)  J &=& 0~.
\eea
\end{subequations}
This is the flavour current supermultiplet \cite{FWZ}.

Let $J^{\a(m) \ad(n)}$ be  a conformal supercurrent of valence $(m,n)$, 
and $\ell_{\a(p) \ad(q)}$ 
a conformal Killing tensor superfield of valence $(p,q)$, 
with $m\geq p$ and $n \geq q$. Then the following composite object 
\bea
{\mathfrak J}^{\a(m-p) \ad(n-q)} [\ell]:= J^{\a(m-p) \b(p) \ad(n-q) \bd (q) } 
\ell_{\b(p) \bd (q)} 
\eea
proves to be a conformal supercurrent of valence $(m-p,n-q)$.

%%%%%%%%%%%%%%%%%%%%%%%%%%%%%%%%
%%%%%%%%%%%%%%%%%%%%%%%%%%%%%%%%

\subsection{Conformal Killing spinor superfields and hypermultiplet} \label{section4.3}

A free superconformal hypermultiplet may be described by 
two primary  superfields of dimension $+1$, 
a chiral scalar $\f$  and a complex linear scalar $\G$, 
\bea
(\bar \cD^2 - 4R ) \G =0~, \qquad {\mathbb A}\G = -\frac 23 \G~.
\label{4.8}
\eea
The corresponding action
\bea
S_{\rm hypermultiplet} = \int \rd^4x\rd^2\q\rd^2\qb\, E \,\Big\{ 
\bar \f \f - \bar \G \G\Big\} ~,
\label{4.9}
\eea
is super-Weyl invariant.\footnote{A chiral scalar $\f$ and a complex linear scalar $\G$ are the physical $\cN=1$ superfields of the arctic hypermultiplet \cite{LR} realised in $\cN=1$ Minkowski superspace.  In addition to $\f$ and $\G$, this off-shell hypermultiplet includes an infinite tail of auxiliary   $\cN=1$ superfields which are complex unconstrained and vanish on-shell. The superconformal arctic hypermultiplets were formulated in \cite{K2006,K2007}. General couplings of arctic hypermultiplets
to 5D $\cN=1$ and 4D $\cN=2$ conformal supergravities were 
presented in \cite{KT-M08-2,KLRT-M}.}

A comment is required
regarding the  $\sU(1)_{R}$ charge assignment 
in \eqref{4.8}.
In general, given a primary complex linear superfield 
$ \G$ of dimension $\D_{ \G}$ and $\sU(1)_{R}$ charge $q_\G$, 
\bea
(\bar \cD^2 - 4R ) \G =0~, \qquad {\mathbb A}\G = q_\G \G~,
\eea
its charge and dimension are related to each other as 
\bea
q_\G = \frac 23 \D_\G - \frac 43 ~,
\eea
as a consequence of the identity
\bea
\d_\S \big(\bar \cD^2 - 4R\big) &=& \S \big(\bar \cD^2 - 4R\big)
- 4(\bar \cD_\ad \S) \bar \cD^\ad +  4(\bar \cD^\ad \S) \bar \cD^\bd \bar M_{\ad\bd} 
- 3(\bar \cD_\ad \S) \bar \cD^\ad {\mathbb A}  \non \\
&& - \frac 32 (\bar \cD^2 \S) {\mathbb A} - 2 (\bar \cD^2 \S) ~.
\eea
The above properties and relations are similar to those 
derived in \cite{KLT-M11} in the case of three-dimensional $\cN=2$ supergravity. 
The $\sU(1)_{R}$ charge of $\G$ was fixed in \eqref{4.8} in order for the action 
\eqref{4.9} to be super-Weyl invariant. 

Given a conformal Killing spinor superfield $\ell_\a$ 
constrained according to \eqref{4.3}, 
\bea
\cD_{(\a} \ell_{\b)} = 0~, \qquad \bar \cD_\ad \ell_\b =0~,
\label{4.13}
\eea
we associate with it the following transformation 
\begin{subequations}\label{secondSUSY}
\bea
\d \f &=& \bar \ell_\ad \bar \cD^\ad \G + \hf (\bar \cD_\ad \bar \ell^\ad) \G~,
\\
\d \G &=&-\ell^\a \cD_\a \f- \hf (\cD^\a \ell_\a) \f~.
\eea 
\end{subequations}
It may be checked that $\bar \cD_\ad \d \f =0$ and $(\bar \cD^2 - 4R ) \d \G =0$.
It may also be verified that $\d \f$ and $\d \G$ are primary superfields. 
A routine calculation shows that the hypermultiplet action \eqref{4.9}
is invariant under the transformation \eqref{secondSUSY},
which  is a curved superspace extension of that given in \cite{K2007}. 

The massless hypermultiplet model \eqref{4.9} has a dual formulation 
realised in terms of two primary dimension-1 chiral scalars
 $\phi$ and $\psi$. The dual action
\bea
\label{HMdual}
S_{\rm hypermultiplet}^{\rm (dual)} = \int \rd^{4}x \rd^{2} \theta \rd^{2} \bar{\theta}\, E\, \Big\{ \bar{\phi} \phi + \bar{\psi} \psi \Big\} ~,
\eea
is obviously super-Weyl invariant. 
In this dual formulation, the rigid symmetry \eqref{secondSUSY} turns into
\begin{subequations}
\label{secondSUSYdual}
\bea
\d \phi & = & \frac{1}{2} \left( \cDB^{2} - 4 R \right) \left( \bar{\ell} \bar{\psi} \right) ~, \\
\d \psi & = & - \frac{1}{2} \left( \cDB^{2} - 4 R \right) \left( \bar{\ell} \bar{\phi} \right) ~.
\eea
\end{subequations}
Here $\bar \ell$ is the complex conjugate of a prepotential $\ell$ defined by 
\bea
\ell_\a = \cD_\a \ell~, \qquad {\mathbb A} \ell = \frac 23 \ell ~.
\label{4.26}
\eea
Equation \eqref{4.13} guarantees the existence of the prepotential  $\ell$, which is 
 is defined modulo arbitrary shifts
\bea
\ell ~\to ~ \ell + \bar \l ~, \qquad\cD_\a \bar \l =0 ~.
\eea
The scalar $\ell$ is primary and of dimension $-1$.

\subsection{Conformal Killing spinor superfields and nonlinear $\s$-models} \label{section4.4}

Now let us return to the nonlinear $\s$-model
\eqref{N=1sigma-model} and 
assume that its target space is a hyperk\"ahler cone \cite{deWRV}.
This means that (i) it is a hyperk\"ahler manifold 
$(\cN, g_{\m\n}, (J_A)^\m{}_\n )$, where $\m,\n=1,\dots, 4n$ and $A=1,2,3$;
and (ii)  it is a K\"ahler cone with respect to each complex structure. 
We pick one of the complex structures, say $J_3$, and introduce complex 
coordinates $\f^i$ compatible with it. In these coordinates, $J_3$ has the form 
\bea
J_3 = \left(
\begin{array}{cc}
{\rm i} \, \d^i{}_j  ~ & ~ 0 \\
0 ~ &   -{\rm i} \, \d^{\bar i}{}_{\bar j}  
\end{array}
\right)~.
\label{J3}
\eea
  Two other complex structures, $J_1$ and $J_2$, become
\bea
J_1 = \left(
\begin{array}{cc}
0  ~ & ~ {g}^{i \bar k} {\bar \o}_{\bar k \bar j} \\
{g}^{ \bar i k } { \o}_{ k  j}
 ~ &   0
\end{array}
\right)~, 
\qquad 
J_2 = \left(
\begin{array}{cc}
0  ~ &  {\rm i}\,   {g}^{i \bar k} {\bar \o}_{\bar k \bar j}  \\
-{\rm i}\,   {g}^{ \bar i k } { \o}_{ k  j}~ &   0
\end{array}
\right)~,
\label{J1J2-hkc}
\eea
where ${g}_{i\bar j} (\f, \bar \f)$ is the K\"ahler metric, and 
$\o_{ij}(\f) = - \o_{ji} (\f)$  is the holomorphic symplectic two-form. 

It may be shown that the $\s$-model action \eqref{N=1sigma-model}
is invariant under the transformation
\bea 
\d \f^i &=&\hf
\big({\bar \cD}^2 -4R \big) \Big\{ \bar{\ell} \,  { \o}^{ij} \c_j \Big\} ~.
\label{hkc-esc}
\eea
The proof is analogous to that given in \cite{K2009} in the case of Minkowski 
superspace.\footnote{See also the seminal paper \cite{HKLR} for the non-superconformal case.} 
If we replace in the right-hand side of \eqref{hkc-esc}  $ \bar{\ell}  \to  \bar{\ell} + \l$, with $\l$ chiral, 
then the $\l$-dependent part of the transformation  a trivial symmetry 
(i.e., it vanishes on-shell) of the model.

%%%%%%%%%%%%%%%%%%%%%%%%%%%%
%%%%%%%%%%%%%%%%%%%%%%%%%%%%%

\subsection{Symmetries of the massless Wess-Zumino operator} \label{section4.5}

Higher symmetries of relativistic wave equations have been  studied over several decades. In particular, it was shown by Shapovalov and Shirokov \cite{ShSh}
and, a decade later,  by Eastwood \cite{Eastwood} that the symmetry algebra of the d'Alembertian 
on ${\mathbb R}^{p,q}$, with $p+q \geq 3$, is isomorphic to (a quotient of) the universal enveloping algebra of the Lie algebra of conformal motions 
that span $\sSO (p+1, q+1)$.  
Such infinite-dimensional algebras and their supersymmetric extensions
play a fundamental role in higher-spin gauge theory \cite{Vasiliev2003}.
Time has come to understand the higher symmetries of supersymmetric extensions of the d'Alembertian. To the best of our knowledge, so far there has appeared only one 
work on the topic, written by Howe and Lindstr\"om \cite{HL2}.

In our discussion of superconformal field theories we re-derived the well-known result that conformal Killing supervector fields generate symmetries of the theories in question \eqref{3.28}.  In this section our analysis will be restricted to the free,  massless theory
obtained from \eqref{3.29} by setting $\l = 0$,
\bea
S[\f, \bar \f] &=&\int \rd^4x\rd^2\q\rd^2\qb  \,E \,\bar \f \f ~.
\eea
 This model  proves to have higher symmetries.
The corresponding super-Weyl invariant equation of motion for $\bar \phi$ is 
\bea
\label{EoMmasslessWZ}
\P \phi  =0~, \qquad \P := -\frac 14 \left( \cD^{2} - 4 \bar{R} \right)~.
\eea
We will refer to $\P$ and its conjugate 
$ \bar \P = -\frac 14 \big( \bar \cD^{2} - 4 {R} \big)$
as the (massless) Wess-Zumino operators.

Here we will study symmetries of the Wess-Zumino operator $\P$. 
A scalar differential operator $\mathfrak O$ will be called
a symmetry operator of $\P$ if it
obeys the two conditions
\begin{subequations}\label{phiConditions}
\bea
\bar \cD_\ad {\mathfrak O} \f &=& 0~, \label{phiConditions-a} \\ 
\P {\mathfrak O} \f &=&0~, \label{phiConditions-b}
\eea
\end{subequations}
for every on-shell chiral scalar $\f$,  \eqref{EoMmasslessWZ}. 
Similar to the non-supersymmetric case \cite{Eastwood}, 
two symmetry 
operators ${\mathfrak O}$ and $\widetilde {\mathfrak O}$ are said to be equivalent, 
$ {\mathfrak O} \sim \widetilde{\mathfrak O}$, if 
\bea
\widetilde{\mathfrak O} - {\mathfrak O} = {\mathfrak F}_\ad \bar \cD^\ad 
+ {\mathfrak H}\P \quad \Longleftrightarrow 
\quad {\mathfrak O} \sim \widetilde{\mathfrak O}~,
\label{equivalence}
\eea
for some operators ${\mathfrak F}_\ad $ and ${\mathfrak H}$.

Since $\f$ is a primary 
superfield of dimension $+1$, we will impose one more condition on $ {\mathfrak O} $, 
which is
\bea
\d_\S ( {\mathfrak O} \f) = \S  {\mathfrak O} \f~.
\label{4.33}
\eea
In other words, we require $\mathfrak O$ to be a conformally invariant operator.
In what follows, we will use  bold-face capital letters, e.g.  $\bm{\mathfrak O}$, 
to denote symmetry operators which only satisfy 
conditions \eqref{phiConditions-a} and \eqref{phiConditions-b}.

Given a positive integer $n$, we look for an $n$th-order  symmetry operator 
\begin{subequations}
\bea
\label{bottomUp}
{\mathfrak O}^{(n)} 
& = & \sum_{k=0}^{n} \z^{A_{1} \dots A_{k}} \cD_{A_{k}} \dots \cD_{A_{1}}  ~,
 \eea
 where the coefficients may be chosen to be graded symmetric 
\bea
   \z^{A_{1} \dots A_{i} A_{i+1} \dots A_{k}} = (-1)^{\ve_{A_{i}} \ve_{A_{i+1}}} \z^{A_{1} \dots A_{i+1} A_{i} \dots A_{k}} ~, \qquad 1\leq i \leq k-1~.
\eea
\end{subequations}
Modulo the equivalence \eqref{equivalence},
${\mathfrak O}^{(n)} $ may be brought to a canonical form given by
\bea
\label{canonicalForm}
{\mathfrak O}^{(n)} = \sum_{k=0}^{n} \z^{\a(k) \ad(k)} \cD_{\a_{1} \ad_{1}} \dots \cD_{\a_{k} \ad_{k}}  + \sum_{k=0}^{n-1} \z^{\a(k+1) \ad(k)} \cD_{\a_{1} \ad_{1}} \dots \cD_{\a_{k} \ad_{k}} \cD_{\a_{k+1}}  ~.
\eea

Now, imposing the condition \eqref{phiConditions-a} proves to lead 
to a number of constraints on the coefficients in   \eqref{canonicalForm}, 
including the following
\label{topComponents}
\bea
\cDB_{\bd} \z^{\a(n) \ad(n)} & = & - 2 \ri \z^{\a(n) ( \ad_{1} \dots \ad_{n-1} } \d^{\ad_{n})}{}_{\bd}
\label{topComponents-a} ~, 
\eea
which is equivalent to
\begin{subequations} \label{4.35}
\bea
\bar \cD_{(\ad_1} \z_{\a(n) \ad_2 \dots \ad_{n+1} ) } &=& 0~, \label{4.35a} \\
\z^{\a(n) \ad(n-1)} & = & \frac{ \ri n}{2(n + 1)} \cDB_{\bd} \z^{\a(n) \ad(n-1) \bd} ~.
 \label{4.35b}
\eea
\end{subequations}
We see that $\z_{\a(n) \ad(n-1)} $ is determined in terms of $\z_{\a(n) \ad(n)} $, and the latter 
is longitudinal linear.
In fact, imposing the condition \eqref{phiConditions-a} also leads to the equation
\bea
\cDB_{\bd} \z^{\a(n) \ad(n-1)} & = & \ri n \z^{\a(n)}{}_{\bd}{}^{\ad(n-1)} R ~,
\eea
which automatically holds as a consequence of \eqref{4.35b}.

Requiring the fulfilment of \eqref{4.33}, 
a routine calculation allows us to express $\z_{\a(n-1) \ad(n-1)}$ in terms of the top component
$\z_{\a(n) \ad(n)}$ as follows:
\bea
\z_{\a(n-1) \ad(n-1)} & = & \frac{n^{2}}{2(n+1)} \cD^{\b \bd} \z_{\b \a(n-1) \bd \ad(n-1)} - \frac{ \ri n^{2}}{4(n+1)(2n+1)} [ \cD^{\b} , \cDB^{\bd} ] \z_{\b \a(n-1) \bd \ad(n-1)} \non \\
& & + \frac{ \ri n (n+1)}{2(2n+1)} G^{\b \bd} \z_{\b \a(n-1) \bd \ad(n-1)} ~.
\label{n-1}
\eea

It should be remarked that the general solution to the constraint  \eqref{4.35a} is
\bea
\z_{\a(n) \ad(n)} = \cDB_{(\ad_{1}} \u_{\a(n) \ad_{2} \dots \ad_{n})} ~, \qquad
{\mathbb A} \u_{\a(n) \ad(n-1)} = - \u_{\a(n) \ad(n-1)} ~,
\label{prepotential}
\eea
where the prepotential $\u_{\a(n) \ad(n-1)} $ is defined modulo 
arbitrary shifts of the form
\bea
\u_{\a(n) \ad(n-1)} \to \u_{\a(n) \ad(n-1)} + \t_{\a(n) \ad(n-1)}~, 
\qquad \bar \cD_{(\ad_1} \t_{\a(n) \ad_2 \dots \ad_{n} ) } =0~.
\eea
Prepotential solution \eqref{prepotential} will be important for our subsequent analysis. 

Suppose we have satisfied \eqref{phiConditions-a}.
Then  imposing the condition \eqref{phiConditions-b} leads to new 
constraints on the coefficients in   \eqref{canonicalForm}, 
including the following
\bea
\cD_{(\a_1} \z_{\a_2 \dots \a_{n+1} ) \ad(n)} = 0~.
\label{4.38}
\eea
Equations \eqref{4.35a} and \eqref{4.38} tell us that the top component $\z_{\a(n) \ad(n)} $
in  \eqref{canonicalForm}  obeys the same constraints \eqref{4.3} which are imposed on the conformal Killing tensor superfield $\ell_{\a(n) \ad(n)}$.
  These constraints are consistent with the reality condition 
  $\bar \z_{\a(n) \ad(n)}  =\z_{\a(n) \ad(n)} $, which will be assumed in what follows. 

So far we have not attempted to find a general solution of the constrains 
\eqref{phiConditions} for ${\mathfrak O}^{(n)} $. Such a solution is easy to work out 
in the case of Minkowski superspace for which 
  a consistent ansatz for an irreducible  
operator 
$\bm{\mathfrak O}^{(n)}$
is given by
\bea
\bm{\mathfrak O}^{(n)}
= \z^{\a(n) \ad(n)} \partial_{\a_{1} \ad_{1}} \dots \partial_{\a_{n} \ad_{n}}  + \z^{\a(n) \ad(n-1)} \partial_{\a_{1} \ad_{1}} \dots \partial_{\a_{n-1} \ad_{n-1}} D_{\a_{n}}  ~,
\label{flatsolution}
\eea
where $D_{A} = ( \partial_{a} , D_{\a} , \bar{D}^{\ad} )$ are the flat superspace covariant derivatives. 
In this case the constraints  \eqref{phiConditions} are equivalent to the relations
\begin{subequations} 
\bea
&&\bar D_{(\ad_1} \z_{\a(n) \ad_2 \dots \ad_{n+1} ) } = 0~, \qquad 
D_{(\a_1} \z_{\a_2 \dots \a_{n+1} ) \ad(n)} = 0~, \\
&&\z_{\a(n) \ad(n-1)} =  - \frac{ \ri n}{2(n + 1)} \bar D^{\bd} \z_{\a(n) \ad(n-1) \bd} ~.
\eea
\end{subequations}
We emphasise that \eqref{flatsolution} is a flat-superspace solution of the constrains 
\eqref{phiConditions}. If the equation \eqref{4.33} is also required, then  
certain lower-order terms must be added to  \eqref{flatsolution}, as follows from 
from eq. \eqref{n-1} and also from the explicit expressions 
for ${\mathfrak O}^{(1)} $ and ${\mathfrak O}^{(2)} $ given below.

The explicit structure of the flat-superspace symmetry \eqref{flatsolution} 
tells us that it is always possible to construct a solution for
the coefficients $\z^{\a(k) \ad(k)} $ and $\z^{\a(k+1) \ad(k)}$ 
of all operators in \eqref{canonicalForm} of order $n-1, \dots ,0$ 
which are proportional to certain components of the torsion tensor and their covariant derivatives. 

The above consideration can be extended to anti-de Sitter superspace, 
${\rm AdS}^{4|4}$,
 which is characterised by the following algebra of covariant derivatives \cite{BK}
 \begin{subequations}  \label{AdSsuperspace}
\bea
&& \qquad \{ \cD_\a , \bar \cD_\ad \} = -2\ri \cD_{\a \ad} ~, \\
&& \qquad \{\cD_\a, \cD_\b \} = -4\bar \m\, M_{\a \b}~, \qquad
\{ {\bar \cD}_\ad, {\bar \cD}_\bd \} = 4\m\,\bar M_{\ad \bd}~, \\
&& \qquad [ \cD_\a , \cD_{ \b \bd }] 
={\rm i} \bar \m\,\ve_{\a \b} \bar \cD_\bd~,  \qquad
\,\,[{ \bar \cD}_{\ad} , \cD_{ \b \bd }] 
=-{\rm i} \m\,\ve_{\ad \bd} \cD_\b~,    \\
&&\quad \,\,\,\,[ \cD_{\a \ad} , \cD_{ \b \bd } ] = -2 \bar \m \m \big({\ve}_{\a \b} \bar M_{\ad \bd }+ \ve_{\ad \bd} M_{\a \b}\big)~,  
\eea
\end{subequations} 
with $\m\neq 0$ being a  complex parameter (the scalar curvature of  
AdS${}_4$ is  equal to $-12 |\m|^2$).
One can show that the following irreducible operator is a consistent ansatz for 
$\bm{\mathfrak O}^{(n)}$ 
\begin{subequations}\label{AdSsolution}
\bea
\bm{\mathfrak O}^{(n)}
= \z^{\a(n) \ad(n)} \cD_{\a_{1} \ad_{1}} \dots \cD_{\a_{n} \ad_{n}}  + \z^{\a(n) \ad(n-1)} \cD_{\a_{1} \ad_{1}} \dots \cD_{\a_{n-1} \ad_{n-1}} \cD_{\a_{n}}  ~.
\eea
Here, the constraints \eqref{phiConditions} are equivalent to
\bea
&&\cDB_{(\ad_1} \z_{\a(n) \ad_2 \dots \ad_{n+1} ) } = 0~, \qquad 
\cD_{(\a_1} \z_{\a_2 \dots \a_{n+1} ) \ad(n)} = 0~, \\
&&\z_{\a(n) \ad(n-1)} =  - \frac{ \ri n}{2(n + 1)} \cDB^{\bd} \z_{\a(n) \ad(n-1) \bd} ~, \\
&&\cDB_{\bd} \z_{\a(n) \a(n-1)} = \ri n \m \z_{\a(n) \ad(n-1)\bd} ~.
\eea
\end{subequations}
We again emphasise that \eqref{AdSsolution} is an AdS-superspace solution of the constraints \eqref{phiConditions}.

We now determine ${\mathfrak O}^{(1)} $ and ${\mathfrak O}^{(2)} $ in $\sU(1)$ superspace.
Setting $n=1$ in 
\eqref{canonicalForm} gives
\bea
{\mathfrak O}^{(1)} \phi & = & \left( \z^{\a \ad} \cD_{\a \ad} + \z^{\a} \cD_{\a} + \z \right) \phi ~.
\label{4.43}
\eea
Requiring  ${\mathfrak O}^{(1)} \phi$ to be chiral allows us to  obtain
\bea
\cDB_{\ad} \z & = & - \frac{\ri}{3} \z^{\a}{}_{\ad} X_{\a} ~.
\label{FirstOrderScalar}
\eea
Additionally, the property that the transformed field remains primary with dimension $+1$ leads to the following super-Weyl transformation laws for the parameters
\begin{subequations}
\label{FirstOrderParametersSW}
\bea
\d_{\S} \z_{\a \ad} & = & - \S \z_{\a \ad} ~, \\
\d_{\S} \z_{\a} & = & - \frac{\S}{2} \z_{\a} - \ri \cDB^{\ad} \S \z_{\a \ad} ~, \\
\d_{\S} \z & = & - 2 \cD^{\a} \S \z_{\a} - \cD^{\a \ad} \S \z_{\a \ad}  + \frac{\ri}{2} [ \cD^{\a} , \cDB^{\ad} ] \S \z_{\a \ad} \label{FirstOrderParametersSW-c}~.
\eea
\end{subequations}
%The transformation laws for the vector and spinor parameters are a consequence of their being a conformal Killing tensor and its descendant, respectively. 
A solution to \eqref{FirstOrderScalar} which is consistent with \eqref{FirstOrderParametersSW-c} is given by
\bea
\z = - \frac{\ri}{3} \u^{\a} X_{\a} 
+ \frac{\ri}{12} \left( \cDB^{2} - 4 R \right) \cD^{\a} \u_{\a} ~,
\label{4.46}
\eea
with the prepotential $\u_\a$ being defined according to \eqref{prepotential}.

It should be emphasised that the second (chiral) term in \eqref{4.46} is not determined by the condition \eqref{phiConditions-a} which only constrains $\z$ to satisfy  \eqref{FirstOrderScalar}.
However, this term is uniquely fixed if we further require the condition \eqref{4.33} to hold.
Making use of the identity
\bea
\left[ \bar \cD^{2}  , \cD_{\a} \right] & = & - 4 \left( G_{\a \ad} - \ri \cD_{\a \ad} \right) 
\bar  \cD^{\ad} 
+4 {R} \cD_{\a}
- 4 \bar \cD^{\ad} G_{\a}{}^{\bd} \bar M_{\ad \bd} + 8 {W}_{\a}{}^{\b \g} {M}_{\b \g} \non \\
&& - \frac{4}{3} {X}^{\b} {M}_{\a \b} - 2 { X}_{\a} \mathbb{A} ~,
\eea
one may obtain from \eqref{4.46} a different expression for $\z$ given by 
\bea
\z = \Big( \frac{1}{4} \cD_{\a \ad} + \frac{\ri}{3} G_{\a \ad} - \frac{\ri}{24} [ \cD_{\a} , \cDB_{\ad} ] \Big) \z^{\a \ad}~,
\eea
which reveals that all parameters of the  operator ${\mathfrak O}^{(1)}$ are expressible in terms of the vector $\z^{\a\ad}$. This is in agreement with the results of the top-down approach \eqref{type1constraints}.

Once the background superspace $(\mathcal{M}^{4|4},\cD)$
possesses first-order symmetry operators 
${\mathfrak O}^{(1)}_{\z_1}, \dots ,  {\mathfrak O}^{(1)}_{\z_n}, $, we can  
generate a higher-order symmetry operator, $\widetilde{\mathfrak O}^{(n)}$,  defined by 
\bea
\widetilde{\mathfrak O}^{(n)}:= 
{\mathfrak O}^{(1)}_{\z_1}  \dots {\mathfrak O}^{(1)}_{\z_n}
~, \qquad n = 2,3, \dots ~.
\eea
By construction, it satisfies the conditions  \eqref{phiConditions} and \eqref{4.33}.
Of course, it does not have the canonical form \eqref{canonicalForm}, however it may be brought to such a form by factoring out a contribution of the type \eqref{equivalence}.

Next, we consider the $n=2$ case
\bea
{\mathfrak O}^{(2)} \phi & = & \left( \z^{\a \b \ad \bd} \cD_{\a \ad} \cD_{\b \bd} + \z^{\a \b \ad} \cD_{\a \ad} \cD_{\b} + \z^{\a \ad} \cD_{\a \ad} + \z^{\a} \cD_{\a} + \z \right) \phi ~.
\eea
Requiring the conditions \eqref{phiConditions} 
leads to the integrability conditions
\begin{subequations}
\bea
\cDB_{\ad} \z_{\b \bd} & = & 2 \ri \ve_{\ad \bd} \z_{\b} + 2 \ri \z^{\a \g}{}_{\ad \bd} W_{\a \b \g} - \ri \z^{\a}{}_{\b \ad \bd} \Big( \cD_{\a} R + \frac{1}{3} X_{\a} \Big) - \ri \z_{\b}{}^{\g}{}_{\ad}{}^{\gd} \cDB_{\gd} G_{\g \bd} \label{integrability-a} \non \\
&& - \z^{\a}{}_{\b (\ad} G_{\a \bd)} ~, \\ 
\cDB_{\ad} \z_{\a}  & = & \ri \z_{\a \ad} R 
+ \frac{\ri}{2} \z_{\a}{}^{\b}{}_{\ad} \Big( \cD_{\b} R + \frac{1}{3} X_{\b} \Big) 
+ \ri \z_{\a}{}^{\b}{}_{\ad}{}^{\bd} \cD_{\b \bd} R - \ri \z^{\b \g}{}_{\ad} W_{\a \b \g} \non \\
&& - \z_{\a}{}^{\b}{}_{\ad}{}^{\bd} G_{\b \bd} ~, \label{integrability-b} \\
\cDB_{\ad} \z & = & \frac{\ri}{3} \z^{\a}{}_{\ad} X_{\a} + \frac{\ri}{3} \z^{\a \b}{}_{\ad} \cD_{\a} X_{\b} + \frac{1}{3} \z^{\a \b}{}_{\ad}{}^{\bd} X_{\a} G_{\b \bd} - \frac{\ri}{3} \z^{\a \b}{}_{\ad}{}^{\bd} \cD_{\b \bd} X_{\a} ~. \label{integrability-c}
\eea
\end{subequations}

So far we have not taken into account the condition \eqref{4.33};
the transformed field, ${\mathfrak O}^{(2)} \phi $,
 retains the property of being primary and of  dimension $+1$.
This condition fixes the super-Weyl transformation laws for the parameters
\begin{subequations}
\bea
\d_{\S} \z_{\a \b \ad \bd} & = & - 2 \S \z_{\a \b \ad \bd} ~, \\
\d_{\S} \z_{\a \b \ad} & = & - \frac{3}{2} \S \z_{\a \b \ad} + 2 \ri \cDB^{\bd} \S \z_{\a \b \ad \bd} ~, \\
\d_{\S} \z_{\a \ad} & = & - \S \z_{\a \ad} - 4 \cD^{\b \bd} \S \z_{\a \b \ad \bd} + \ri [ \cD^{\b} , \cDB^{\bd} ] \S \z_{\a \b \ad \bd} - 4 \cD^{\b} \S \z_{\a \b \ad} ~, \\
\d_{\S} \z_{\a} & = & - \frac{1}{2} \S \z_{\a} + \ri \cDB^{\ad} \cD^{\b \bd} \S \z_{\a \b \ad \bd} - 2 \cD^{\b \bd} \S \z_{\a \b \ad \bd} + \ri \cD^{\ad} \S \z_{\a \ad} ~, \\
\d_{\S} \z & = & \frac{\ri}{2} \cD^{\a \ad} [ \cD^{\b} , \cDB^{\bd} ] \S \z_{\a \b \ad \bd} - 2 \cD^{\a} \cD^{\b \bd} \S \z_{\a \b \bd} + \frac{\ri}{2} [ \cD^{\a} , \cDB^{\ad} ] \S \z_{\a \ad} \non \\
&& - 2 \cD^{\a} \S \z_{\a} - \cD^{\a \ad} \S \z_{\a \ad} - \cD^{\a \ad} \cD^{\b \bd} \S \z_{\a \b \ad \bd} ~.
\eea
\end{subequations}

The requirement that \eqref{phiConditions} and \eqref{4.33} are satisfied leads to the unique solution
\begin{subequations}
\bea
\z_{\a \ad} & = & \frac{2}{3} \cD^{\b \bd} \z_{\a \b \ad \bd} - \frac{\ri}{15} [ \cD^{\b} , \cDB^{\bd} ] \z_{\a \b \ad \bd} + \frac{3 \ri }{5} \z_{\a \b \ad \bd} G^{\b \bd} ~, \\
\z_{\a} & = & - \frac{2 \ri}{15} \cDB^{\ad} \cD^{\b \bd} \z_{\a \b \ad \bd} - \frac{1}{10} \z_{\a \b \ad \bd} \cDB^{\ad} G^{\b \bd} + \frac{1}{15} \cDB^{\ad} \z_{\a \b \ad \bd} G^{\b \bd} ~, \\
\z & = & \frac{1}{15} \cD^{\a \ad} \cD^{\b \bd} \z_{\a \b \ad \bd} - \frac{\ri}{60} \cD^{\a \ad} [ \cD^{\b} , \cDB^{\bd} ] \z_{\a \b \ad \bd} + \frac{7 \ri}{30} \cD^{\a \ad} \z_{\a \b \ad \bd} G^{\b \bd} \non \\
&& + \frac{1}{30} [ \cD^{\a} , \cDB^{\ad} ] \z_{\a \b \ad \bd} G^{\b \bd} + \frac{1}{20} \cD^{\a} \z_{\a \b \ad \bd} \cDB^{\ad} G^{\b \bd} - \frac{13}{60} \cDB^{\ad} \z_{\a \b  \ad \bd} \cD^{\a} G^{\b \bd} \non \\
&& + \frac{2 \ri}{5} \z_{\a \b \ad \bd} \cD^{\a \ad} G^{\b \bd} + \frac{3}{20} \z_{\a \b \ad \bd} [ \cD^{\ad} , \cDB^{\ad} ] G^{\b \bd} - \frac{1}{5} \z_{\a \b \ad \bd} G^{\a \ad} G^{\b \bd} ~.
\eea
\end{subequations}
Thus, this transformation is completely determined by the conformal Killing tensor $\z^{\a \b \ad \bd}$. It is crucial to note that if we relax condition \eqref{4.33}, the solution ceases to be uniquely defined and may be constructed in such a way that the coefficients $\z_{\a \ad} ~,~ \z_{\a}$ and $\z$ vanish in the flat (or AdS) superspace limit.

In the case of a symmetry operator \eqref{canonicalForm} of arbitrary order, we expect that our conceptual results for ${\mathfrak O}^{(1)} $ and ${\mathfrak O}^{(2)} $
 generalise; all components are uniquely determined in terms of $\z_{\a(n) \ad(n)}$ and a suitable flat (or AdS) superspace limit may be constructed.

%%%%%%%%%%%%%%%%%%%%%%%%%%%%%
%%%%%%%%%%%%%%%%%%%%%%%%%%%%%

\subsection{Supersymmetric even Schouten-Nijenhuis bracket}

In analogy with the space of conformal Killing supervector fields, we wish to endow our construction with an additional structure allowing us to combine two conformal Killing tensors and produce a third. Consider two such tensors $\z^{1}_{\a(m) \ad(m)}$ and $\z^{2}_{\a(n) \ad(n)}$. It can then be shown that 
the following bracket (an implicit symmetrisation over all  $\a$-indices and, independently,
all  $\ad$-indices is assumed below)
\bea
&& [ \z^{1} , \z^{2} ]_{\a(m+n-1) \ad(m+n-1)} =  - \frac{m}{2} \z^{1}_{\a(m-1)}{}^{\b}{}_{\ad(m-1)}{}^{\bd} \cD_{\b \bd } \z^{2}_{\a(n) \ad(n)} + \frac{n}{2} \z^{2}_{\a(n-1) }{}^{\b}{}_{\ad(n-1)} {}^{\bd} \cD_{\b \bd } \z^{1}_{\a(m) \ad(m)} \non \\
&& - \frac{\ri m n}{4(m+1)(n+1)} \left( \cDB_{\bd} \z^{1}_{\a(m)}{}^{\bd}{}_{\ad(m-1)} \cD_{\b} \z^{2}_{\a(n-1)}{}^{\b}{}_{\ad(n)} - \cDB_{\bd} \z^{2}_{\a(n)}{}^{\bd}{}_{\ad(n-1)} \cD_{\b} \z^{1}_{\a(m-1)}{}^{\b}{}_{\ad(m)} \right) \non \\
&& + \frac{\ri m n}{2} \left( \z^{1}_{\a(m) \ad(m-1)}{}^{\bd} \z^{2}_{\a(n-1)}{}^{\b}{}_{\ad(n-1)}- \z^{2}_{\a(n) \ad(n-1)}{}^{\bd} \z^{1}_{\a(m-1)}{}^{\b}{}_{\ad(m-1)} \right) G_{\b \bd}
\label{4.56}
\eea
 also satisfies these conditions and hence is a new conformal Killing tensor superfield.
 Hence, for a given supergravity background, 
the set of conformal Killing tensor superfields $\z_{\a(n) \ad(n)}$
is a superalgebra with respect to the above bracket.
 
 The $G_{\b\bd}$-dependent terms in \eqref{4.56} can be removed by redefining the vector covariant derivative by the rule
 \begin{subequations}
 \bea
 \cD_{\a\ad} ~\to ~\widetilde{\cD}_{\a\ad} = \cD_{\a\ad} 
 + \frac{\ri}{2} \big( G^\b{}_\ad M_{\a\b} - G_\a{}^\bd \bar M_{\ad \bd} \Big)
 \eea
 or, equivalently, 
 \bea
 \cD_a ~\to ~\widetilde{\cD}_a + \frac 14 \ve_{abcd} G^b M^{cd}~.
 \eea
\end{subequations}
The specific  feature of the covariant derivatives 
$\widetilde{\cD}_A = (\widetilde{\cD}_a, \cD_\a , \bar \cD^\ad)$ is the torsion-free condition
$\widetilde{T}_{ab}{}^c =0$. In terms of the covariant derivatives $\widetilde{\cD}_A$, 
the bracket \eqref{4.56} coincides with the one proposed in \cite{HL1}
where it was called the ``supersymmetric even Schouten-Nijenhuis bracket.''

In the case of $\cN=1$ AdS superspace, the bracket \eqref{4.56} coincides with the one given 
in \cite{GKS} for Killing tensor superfields. 

%%%%%%%%%%%%%%%%%%%%%%%%%%%%
%%%%%%%%%%%%%%%%%%%%%%%%%%%%

\section{Isometries of curved superspace}\label{section5}

As is well known, every off-shell formulation for $\cN=1$ supergravity 
is obtained by coupling conformal supergravity to a compensating supermultiplet. 
Different supergravity theories correspond to different  compensators,
see, e.g., \cite{SG,deWR,FGKV,GGRS}. 
For a given theory, the compensator $\X$ is a nowhere vanishing 
primary scalar superfield, which obeys certain constraints and has 
a non-zero dimension, $\D_\X\neq 0$, and some $\sU(1)_R$ charge $q_\X$. 
In the case of new minimal supergravity,  $q_\X= 0$ and the compensator is real. 
For the old minimal and non-minimal formulations, $q_\X$ is non-zero.
Once $\X$ is specified, supergravity background is a triple $(\cM^{4 |4}, \cD,\X)$.

\subsection{Off-shell supergravity and Killing vector superfields}

Let $\x= \x^B E_B$ be a conformal Killing supervector field on $(\cM^{4 |4}, \cD)$,
\begin{subequations} \label{Killing}
\bea
\big(\d_{\cK [\x]} + \d_{\S [\x]} \big) \cD_A = 0~.   \label{Killing-a}
\eea
 It is called a Killing supervector field 
 if  it leaves the compensator $\X$ invariant, 
\bea
\big( \d_{\cK [\x]} + \D_\X \S[\x] \big) \X =0~.  \label{Killing-b}
\eea
\end{subequations}
The latter condition can be rewritten in the form 
\bea
\label{CompensatorKilling}
 \xi^{B} \cD_{B}  \X + ( \D_\X \S [\xi] + {\rm i} q_\X \r [\xi]) \X= 0 ~.
\eea
 The set of all Killing supervector fields on $(\cM^{4 |4}, \cD,\X)$
is a Lie superalgebra. 

The Killing equations \eqref{Killing} are super-Weyl invariant in the sense that 
they hold for all conformally related supergravity backgrounds.  In the presence of a compensator,
the notion of conformally related superspaces given in section \ref{section3.3}
should be generalised  as follows.
Two supergravity backgrounds $(\cM^{4 |4}, \hat{\cD},{\hat \X})$ and $(\cM^{4 |4}, \cD,\X)$ 
 are said to be conformally related provided
the covariant derivatives $\hat{\cD}_{A}$
and $\cD_{A}$ are related to each other according to \eqref{ConRel}, and the same super-Weyl parameter $\S$ relates the compensators,
\bea
{\hat \X} = \re^{\D_\X \S} \X~.
\eea

Applying a super-Weyl transformation allows us to choose the gauge
\bea
\bar \X \X=1~,
\label{5.4}
\eea
and then \eqref{CompensatorKilling} reduces to 
\bea
\S[\x] =0 \quad \Longleftrightarrow \quad 
\cD^{\a} \xi_{\a} + \cDB_{\ad} {\bar \xi}^{\ad} = 0 
\quad \implies \quad  \cD_{a} \xi^{a} = 0 ~.
\eea
 In this gauge the Killing equations \eqref{Killing} take the simplified form
\bea
\label{killingSupervector}
\d_{\mathcal{K} [ \xi ] } \cD_{A} = \left[ \mathcal{K} [ \xi ]  , \cD_{A} \right] = 0 ~.
\eea
Once $\S [\x]= 0$ the left-hand side of each relation in \eqref{3.14}
is equal to zero, and therefore the right-hand side must vanish as well.
It is an instructive exercise to demonstrate, with the aid of the relations \eqref{type2constraints1} and \eqref{type2constraints2}, that this is indeed the case.

For $q_\X \neq 0$ it is always possible to impose a stronger gauge condition than \eqref{5.4}. 
Indeed, applying a combined super-Weyl and local $\sU(1)_R$ transformation allows us to set 
\bea
\X =1~,
\label{5.7}
\eea
and then the Killing condition \eqref{CompensatorKilling} turns into  
\bea
\S[\x] =0~, \qquad \xi^{B} \F_{B}   + \r [\xi] = 0 ~.
\eea

When studying the symmetries of bosonic supergravity backgrounds, we will keep some of the components of $\X$ alive and, instead, impose the so-called  Weyl multiplet gauge described in 
Appendix \ref{Weyl_multiplet}.

%%%%%%%%%%%%%%%%%%%%%%%%%%%%%%%%
%%%%%%%%%%%%%%%%%%%%%%%%%%%%%%%%

\subsection{Conformal compensators} 

In this subsection we briefly review the structure of the compensating supermultiplets which correspond to the old minimal \cite{WZ,old1,old2}
and  new minimal \cite{new} formulations for $\cN=1$ supergravity.
The non-minimal formulations for Poincar\'e \cite{non-min,SG} and AdS supergravity
 \cite{ButterK} will not be discussed here.

In the old minimal formulation, the compensator is a nowhere vanishing primary 
chiral scalar $S_0$ with the superconformal properties 
\bea
\bar \cD_\ad S_0 =0~, \qquad \D_{S_0} =1 ~, \qquad q_{S_0} = \frac 23~. 
\label{5.9}
\eea
The supergravity action is
\bea
S_{\text{SG,old}} = -\frac{3}{\k^2} \int {\rm d}^4x \rd^2\q\rd^2\qb
\,E\, \bar S_0  S_0
+ \Big\{ \frac{\m}{\k^2} \int {\rm d}^4x {\rm d}^2 \q \,\cE\,   S_0^3  + {\rm c.c.} \Big\} ~,
\label{old-minimal}
 \eea
where $\k$ is the gravitational coupling constant, and 
$\m$ is a complex parameter related to the cosmological constant.
Making use of the super-Weyl and local $\sU(1)_R$ transformations, the chiral compensator can 
be gauged away resulting with 
\bea
S_0 =1 \quad \implies \quad \F_A=0 \quad \implies \quad X_\a =0~.
\eea

In the new minimal formulation, the compensator is a nowhere vanishing 
primary scalar constrained by\footnote{The linear compensator \eqref{5.12}
was introduced in \cite{deWR}. It is  a tensor multiplet
\cite{Siegel-tensor} such that its field strength $L$ is
nowhere vanishing.} $ L$, 
\bea  
\bar { L}= { L}~, \qquad 
(\bar \cD^2 -4R) { L}  =0 \quad \implies \quad \D_L = 2~.
\label{5.12}
\eea
The supergravity action is 
\bea
S_{\text{SG,new}}= \frac{3}{\k^2} \int \rd^4 x \rd^2 \q  \rd^2 \bar{\q} \, E\,
{L}\, {\rm ln} \frac{L}{\bar S_0 S_0} ~,
\eea
where the chiral scalar $S_0$, eq. \eqref{5.9} is a pure gauge degree of freedom.
The super-Weyl invariance allows one to choose the gauge
\bea
L=1 \quad \implies \quad R=0~.
\eea

%%%%%%%%%%%%%%%%%%%%%%%%%%%
%%%%%%%%%%%%%%%%%%%%%%%%%%%%

\subsection{Killing spinor superfield and  massive hypermultiplet}

To describe a massive hypermultiplet, we consider the following generalisation of \eqref{HMdual}
\bea
S_{\rm hypermultiplet}^{(m)} 
= \int \rd^{4}x \rd^{2} \theta \rd^{2} \bar{\theta} \, E\, \Big\{ \bar{\phi} \phi + \bar{\psi} \psi \Big\} + \Big\{m\,  \ri   \int\rd^4x\rd^2\q\, \cE \,S_0 \psi \f +{\rm c.c.} \Big\} ~,
\eea
where $m$ is a real mass parameter. When analysing this model, we will adopt the super-Weyl gauge $S_0 = 1$, and therefore the $\sU(1)_R$ connection is equal to zero, $\F_A=0$. 

Through a direct computation, we find that the transformation \eqref{secondSUSYdual} is also a symmetry of the massive theory only if  $\ell$ is constrained to be real, 
\begin{subequations}
\bea
\bar \ell = \ell 
\quad \implies \quad \cD_{\a\ad} \ell =0~,
\eea
where we have used the relations
\eqref{4.13} and \eqref{4.26}, which imply 
\bea
\bar \cD_\ad \cD_\a \ell =0 ~.
\eea
\end{subequations}
 These conditions may be shown to have the following non-trivial implication:
\bea
\cD_{\a} \left( \cDB^{2} - 4 R \right) \ell = - 4 G_{\a \ad} \cDB^{\ad} \ell ~.
\eea
Now, in conjunction with the identity $\bar \cD_{\ad} \left( \cDB^{2} - 4 R \right) \ell =0$, 
we observe that 
\bea
G_{\a\ad}=0 \quad \implies \quad  \left( \cDB^{2} - 4 R \right) \ell = {\rm const} ~.
\eea
The condition $G_{\a\ad}=0$ means that the background under consideration is Einstein, i.e. it is a solution of supergravity equations of motion. 

To realise a second supersymmetry transformation in $\cN=1$ AdS superspace, 
Refs.  \cite{GKS-susy,GKS} made use of
a background scalar superfield $\ve$ subject to the constraints
\begin{align}
\bar \ve =\ve~, \qquad 
 \bar\cD_\ad \cD_\alpha \ve = 0~, \qquad (\bar \cD^2 - 4 \mu) \ve =0 ~.
\label{epsilon-constraints}
\end{align}
The parameter $\ve$ naturally originates within the $\cN=2$ AdS superspace approach \cite{KT-M08}. The Killing superfield $\ell$ introduced above contains two additional scalar parameters as compared with $\ve$.

%%%%%%%%%%%%%%%%%%%%%%%%%%%%%%%%%%%%
%%%%%%%%%%%%%%%%%%%%%%%%%%%

\subsection{Symmetries of the massive Wess-Zumino operator}

A massive scalar supermultiplet in curved superspace 
is described by the action
\bea
S[\f, \bar \f] &=&\int \rd^4x\rd^2\q\rd^2\qb  \,E \,\bar \f \f 
+ \Big\{ \frac{m}{2}  \int\rd^4x\rd^2\q\, \cE \,S_0\f^2 +{\rm c.c.} \Big\}~,
\qquad \bar \cD_\ad \f =0~, ~~
\eea
with $m=\bar m $ a mass parameter.
In what follows we will work in the super-Weyl gauge $S_0=1$. 
Then the equations of motion are
\bea
\cH_m 
\left(\begin{array}{c}\f \\ \bar \f \end{array}\right) =0~,
\qquad
\cH_m=\left(\begin{array}{cc} m~&\bar \P  \\ 
 \P & m
\end{array}\right)~, \qquad \P := -\frac 14 \left( \cD^{2} - 4 \bar{R} \right)~.
\eea

We now wish to understand what additional conditions must be imposed upon the $n$th-order  operator \eqref{canonicalForm} so that we obtain a symmetry of this theory. Since it has been shown that all coefficients
are expressed in terms of the top component, we expect that this condition may be written as a closed
form equation in $\z_{\a(n) \ad(n)}$.

In the massive case, the requirement that the symmetry operator $\mathfrak{O}^{(n)}$ preserves the equation of motion
\bea
\P \mathfrak{O}^{(n)} \phi + m \overline{ \left( \mathfrak{O}^{(n)} \phi  \right) } = 0 ~,
\eea
leads to new conditions which arise from setting the contributions proportional to the derivatives of $\bar \phi$ to zero. The most fundamental of these is
\bea
\label{5.10}
\cD^{\b} \cDB^{\bd} \z_{\b \a(n-1) \bd \ad(n-1)} & = & 2 ( n + 1 ) G^{\b \bd} \z_{\b \a( n - 1 ) \bd \ad(n - 1)} \non \\
&  &+ \frac{2 \ri (n+1)}{n} \left( \z_{\a(n-1) \ad(n-1)}  - \bar{\z}_{\a(n-1) \ad(n-1)} \right) ~.
\eea

It is more useful to work with an expression only in terms of the top component. 
Substituting \eqref{n-1} into \eqref{5.10} yields the Killing condition
\bea
\label{KillingTensor}
\cD^{\b} \cDB^{\bd} \z_{\b \a(n-1) \bd \ad(n-1)} = 2 n ( n + 1 ) G^{\b \bd} \z_{\b \a( n - 1 ) \bd \ad(n - 1)} ~,
\eea
which implies 
\bea
 \cD^{\b \bd} \z_{\b \a(n-1) \bd \ad(n-1)} = 0 ~.
\eea
Fixing $n = 1$, we obtain the well-known Killing condition for supervector fields \eqref{1.3}.

In the case of AdS superspace AdS${}^{4|4}$, 
$G_{\a \ad} = 0$ and the Killing condition \eqref{KillingTensor} reduces to \eqref{1.4b} 
originally described in \cite{GKS}. Given two Killing tensor superfields
$\z^{1}_{\a(m) \ad(m)}$ and $\z^{2}_{\a(n) \ad(n)}$ in AdS${}^{4|4}$, 
the bracket \eqref{4.56} coincides with the one presented in \cite{GKS}.

%%%%%%%%%%%%%%%%%%%%%%%%%%%%
%%%%%%%%%%%%%%%%%%%%%%%%%%%

\section{Bosonic backgrounds}

Similar to general relativity, of special interest are supergravity backgrounds which support  unbroken symmetries.
In the context of supersymmetric field theory we are primarily interested in those backgrounds which
possess some amount of unbroken supersymmetry. 
This naturally leads us to restrict our attention to so-called  
bosonic backgrounds. By definition such a supergravity background has no covariant fermionic fields, 
\bea
\label{covFermionic}
\cD_{\a} R | = 0 ~, \qquad \cD_{\a} G_{\b \bd} | = 0 ~, \qquad W_{\a \b \g} | = 0 ~, \qquad X_{\a}| = 0 ~,
\eea
where the bar  projection is defined as in eq. \eqref{B.1}.
These conditions imply that the gravitino can be gauged away. In the remainder of this section we will 
assume that the gravitino is absent. We will also make use of the Weyl multiplet gauge 
described in Appendix \ref{Weyl_multiplet}.

Since there are no background fermionic fields, it follows from the equations \eqref{type2constraints1} 
that every conformal Killing supervector field can uniquely be written as a sum of even and odd ones.
A conformal Killing supervector field $\x^A$ is called even if 
\bea
v^{a}(x)  := \xi^{a} | \neq 0 ~, \qquad \xi^{\a} | = 0 ~.
\label{6.2}
\eea
A conformal Killing supervector field $\x^A$ is called odd provided
\bea
\xi^{a} | \, = 0 ~, \qquad \e^{\a} (x):= \xi^{\a}| \, \neq 0 .
\label{6.3}
\eea
All information about the even and odd conformal Killing supervector fields is encoded in the vector 
$v^a$ and spinor $\e^\a$ fields, respectively.

%%%%%%%%%%%%%%%%%%%%%%%%%%
%%%%%%%%%%%%%%%%%%%%%%%%%%

\subsection{Conformal isometries}

In this section 
we make extensive use of the component field formalism reviewed in Appendix \ref{AppendixB} 
and work within the Weyl multiplet gauge constructed in Appendix \ref{Weyl_multiplet}. 
Since the gravitino has been gauged away, which is possible due to \eqref{covFermionic}, the component torsion tensor  \eqref{C.3b} vanishes, 
which leaves us with a torsionless Lorentz connection. The component covariant derivative is
\bea
\cD_{a} | = \mathfrak{D}_{a} ~, \qquad 
\big[ \mathfrak{D}_{a} , \mathfrak{D}_{b} \big] = \frac{1}{2} R_{ab}{}^{cd} M_{cd} + \ri F_{a b} \mathbb{A} ~.
\eea
where the Lorentz curvature and $\sU(1)_{R}$ field strength take the form
\begin{subequations}
\bea
R_{abcd} & = & \frac{1}{2} (\s_{a b})^{\a \b} (\s_{cd})^{\g \d} \cD_{(\a} W_{\b \g \d)} |
- \frac{1}{2} (\tilde{\s}_{ab})^{\ad \bd} (\tilde{\s}_{cd})^{\gd \dd} \cDB_{(\ad} \bar{W}_{\bd \gd \dd)} | \non \\
&  & + \frac{1}{4} \Big( (\tilde{\s}_{a b})^{\ad \bd} (\s_{c d})^{\a \b} + (\s_{a b})^{\a \b} (\tilde{\s}_{cd})^{\ad \bd} \Big) \cD_{(\a} \cDB_{(\ad} G_{\b) \bd)} | \non \\
&  & - \frac{1}{24} \Big( \eta_{c[a} \eta_{b]d} - \eta_{d[a} \eta_{b] c} \Big) \cD^{\a} X_{\a} | ~, \\
F_{ab} & = & \frac{\ri}{8} (\s_{ab})^{\a \b} \cD_{\a} X_{\b} | - \frac{\ri}{8} (\tilde{\s}_{a b})^{\ad \bd} \cDB_{\ad} \bar{X}_{\bd} | ~.
\eea
\end{subequations}
When working with a $\sU(1)_R$ neutral field $\j(x) $, it holds that  $\mathfrak{D}_{a} \j = \nabla_a \j$, 
where
\bea
\nabla_{a} := \mathfrak{D}_{a} - \ri \varphi_{a} \mathbb{A} 
\label{6.6}
\eea
is the torsion-free Lorentz-covariant derivative.

In Section \ref{Section3}, we derived the necessary conditions on the transformation parameters $\bm \X$, 
eq. \eqref{conformalparameters}, associated with a conformal Killing supervector field $\x^A$.
Here, we wish to extract from these conditions all the restrictions on even and odd conformal
Killing supervector fields. These are readily derivable by bar projecting the results for $\cD_{a} \bm \Xi$.

Let $\x^A$ be an even conformal Killing supervector field.
Making use of the definitions \eqref{B.11} and bar projecting eq. \eqref{CKVector}
leads to 
\bea
\label{bosonicKV}
\nabla_{a} v_{b} = k_{a b} [ v ] + \eta_{a b} \s [ v ] ~,
\eea
which implies
\bea
k_{a b} [ v ] = \nabla_{[a} v_{b]} ~, \qquad \s [ v ] = \frac{1}{4} \nabla_{a} v^{a} ~.
\eea
We see that 
$v^{a}$ is a conformal Killing vector field, 
\bea
\nabla_{ ( a}  v_{b )} = \frac{1}{4} \eta_{a b} \nabla_{c} v^{c} ~.
\eea
Further, one may show that every conformal Killing vector field 
on $\cM^4$ may be lifted to a unique even conformal Killing supervector field on $\cM^{4|4}$.
It should be remarked that the $\sU(1)_{R}$ parameter $ \varrho [ v ] $ is given by 
\bea
\nabla_{a} \varrho [ v ] = - F_{ab} v^{b} ~.
\eea

Let $\x^A$ be an odd conformal Killing supervector field, eq. \eqref{6.3}.
Then the bar projection of \eqref{CKSpinor} yields the conformal Killing spinor equation
\bea
\label{fermionicKV}
\mathfrak{D}_{\a \ad} \e_{\b} = - { \rm i} \ve_{\a \b} {\bar \eta}_{\ad} [ \e ] ~,
\eea
where we have defined $ \eta_{\a} := \cD_{\a} \S | $. 
The equivalent form of this equation is
\bea
\mathfrak{D}_{(\a \ad} \e_{\b)} =0~.
\eea

%%%%%%%%%%%%%%%%%%%%%%%%%%%%%%%
%%%%%%%%%%%%%%%%%%%%%%%%%%%%%%%%

\subsection{Isometries of old minimal supergravity backgrounds}

Let $\x= \x^B E_B$ be a conformal Killing supervector field on $(\cM^{4 |4}, \cD)$, 
eq. \eqref{Killing-a}. 
We recall that the transformation $\d_{\cK [\x]} + \d_{\S [\x]}$  is said to be an isometry if the conformal compensator is left invariant, eq.  \eqref{Killing-b}. In general, this requirement leads to severe restrictions on the symmetry parameters. Here, we will investigate the case of old minimal supergravity.

By making use of the Weyl, local  $\sU(1)_{R}$ and S-supersymmetry transformations we are able to adopt the gauge
\bea
S_{0} | \, = 1 ~, \qquad \cD_{\a} S_{0} | \, = 0 ~.
\eea
This leaves us with a single component field which cannot be gauged away
\bea
M : = - \frac{1}{4} \cD^{2} S_{0} | ~.
\eea
As we have fixed the local $\sU(1)_{R}$ invariance in this gauge, it is more convenient 
to work with the Lorentz-covariant derivative \eqref{6.6}.

We find that in the case of an even symmetry, equation \eqref{Killing-b} is equivalent to the conditions
\bea
\s [ v ] = 0 ~, \qquad \varrho [ v ] = v^{a} \varphi_{a} ~, \quad v^{a} \nabla_{a} M = 0 ~.
\eea
As a result, \eqref{bosonicKV} reduces to
\bea
\nabla_{a} v_{b} = k_{a b} [ v ]  \quad \implies \quad \nabla_{(a} v_{b)} = 0 ~,
\eea
and therefore $v^a$ is a Killing vector field.

If we instead consider odd symmetries, we obtain
\bea
\eta_{\a} [ \e ] = - M \e_{\a} - \frac{2}{3} \bar{\e}^{\ad} \varphi_{\a \ad} ~.
\eea
Thus, we are able to obtain from \eqref{fermionicKV} the Killing spinor equation
\bea
\nabla_{\a \ad} \e_{\b} = \ri \varphi_{ ( \a \ad} \e_{\b)} + \ri \ve_{\a \b} \left( \bar{M} \bar{\e}_{\ad} 
+ \frac{1}{6} \varphi_{\g\ad} \e^{\g} \right) ~,
\eea
which was originally given in \cite{FS}.

%%%%%%%%%%%%%%%%%%%%%%%%%
%%%%%%%%%%%%%%%%%%%%%%%%%

\subsection{Isometries of new minimal supergravity backgrounds}

In the case of new minimal supergravity, the conformal compensator $L$ is a 
linear multiplet, eq. \eqref{5.12}. Associated with $L$ is the real vector descendant
\bea
\label{vecCompensator}
L_{\a \ad} := - \frac{1}{2} \left[ \cD_{\a} , \cDB_{\ad} \right] L + G_{\a \ad} L 
\eea
with the important property
\bea
\label{vectorCompensatorfield}
\cD^{\a \ad} L_{\a \ad} & = & \frac{\rm i}{2} \left( {\bar X}_{\ad} + 3 \cDB_{\ad} {\bar R} \right) \cDB^{\ad} L - \frac{\rm i}{2} \left( X^{\a} + 3 \cD^{\a} R \right) \cD_{\a} L ~.
\eea

Working in the Weyl multiplet gauge, the freedom to perform the Weyl and S-supersymmetry transformations allows us to impose the additional gauge conditions
\bea
L | \, = \, 1 ~, \qquad \cD_{\a} L | \, = \, 0 ~.
\eea
Owing to the reality of $L$, we stay with unbroken $\sU(1)_{R}$ transformations. 
The only remaining component field of $L$ is 
\bea
H_{\a \ad} := L_{\a \ad} \, | ~.
\eea
Making use of \eqref{vectorCompensatorfield}, we arrive at the constraint
\bea
\nabla^{a} H_{a} = 0 ~.
\eea

Considering the case of an even symmetry, equation \eqref{Killing-b} leads to
\bea
\s [ v ] = 0 ~, \qquad v^{b} \nabla_{b} H_{a} = 0 ~.
\eea
As a result, the Killing vector equation is given by
\bea
\nabla_{a} v_{b} = k_{a b} [ v ] \quad \implies \quad \nabla_{(a} v_{b)} = 0 ~.
\eea

In the case of odd symmetries, we deduce the charged Killing spinor equation
\bea
\eta_{\a} [ \e ] = -\hf H_{\a\bd} \bar \e^\bd \quad \implies \quad 
 \mathfrak{D}_{\a \ad} \e_{\b} = \frac{\ri}{2} \ve_{\a \b} H_{\g \ad} \e^\g ~,
\eea
which is equivalent to the one originally derived in \cite{FS}.

%%%%%%%%%%%%%%%%%%%%%%%%%%%%
%%%%%%%%%%%%%%%%%%%%%%%%%%%

\subsection{Components of the (conformal) Killing tensor superfields}

Given a primary tensor field $t_{\a(p) \ad(q)}$  on a curved spacetime, we say that it is conformal Killing if it satisfies
\bea
\label{conformalKillingBBG}
\mathfrak{D}_{\a \ad } t_{\a(p) \ad(q)} = 0 ~.
\eea
Further, it is said to be Killing if
\bea
\label{KillingBBG}
\mathfrak{D}^{\b \bd} t_{\b \a(p-1) \bd \a(q - 1)} ~.
\eea

Consider a conformal Killing tensor superfield $\ell_{\a(m) \ad(n)}$ on $\cM^{4|4}$
with $m \geq 1$  and  $n \geq 1$. It obeys the constraints
\eqref{4.3a} and \eqref{4.3b}. 
At the component level it contains four independent fields:
\begin{subequations}
\bea
K_{\a(m) \ad(n)} & : = &  \ell_{\a(m) \ad(n)} | ~, \\
M_{\a(m-1) \ad(n)}  &: =&  \cD^{\b} \ell_{\b \a(m-1) \ad(n)} | ~, \\
N_{\a(m) \ad(n-1)}  &: =&   \cDB^{\bd} \ell_{\a(m) \bd \ad(n-1)} | ~, \\
 L_{\a(m-1) \ad(n-1)}  &: =& [ \cD^{\b} , \cDB^{\bd} ] \ell_{\b \a(m-1) \bd \ad(n-1)} | ~.
\eea
\end{subequations}
By a straightforward calculation, we find that each component field defines a conformal Killing tensor field on the background in the sense of \eqref{conformalKillingBBG}. In the special case where $\ell_{\a(m) \ad(n)}$ is Killing, it is easily shown that these component fields also satisfy the Killing condition \eqref{KillingBBG}.

%%%%%%%%%%%%%%%%%%%%%%%%%%%%%%%%
%%%%%%%%%%%%%%%%%%%%%%%%%%%%%%%%%

\subsection{Components of conformal supercurrents}

A primary tensor field $t^{\a(p) \ad(q)}$ on a curved spacetime 
will be called a conserved current if it satisfies
the divergenceless condition
\bea
\label{ConservedCurrent}
\mathfrak{D}_{\b \bd} t^{\b \a(p-1) \bd \ad(q-1)} = 0 ~.
\eea

Given a conformal supercurrent $J^{\a(m) \ad(n)}$, eq.  \eqref{supercurrent}, it contains four independent component fields, which can be chosen as follows
(an implicit symmetrisation over all  $\a$-indices and, independently,
all  $\ad$-indices is assumed)
\begin{subequations}
\bea
j^{\a(m) \ad(n)} & := & J^{\a(m) \ad(n)} | ~, \\
Q^{\a(m+1) \ad(n)} & := & \cD^{\a} J^{\a(m) \ad(n)} | ~, \\
S^{\a(m) \ad(n+1)} & := & \cDB^{\ad} J^{\a(m) \ad(n)} | ~, \\
T^{\a(m+1) \ad(n+1)} & := & [ \cD^{\a} , \cDB^{\ad} ] J^{\a(m) \ad(n)} | ~.
\label{6.31d}
\eea
\end{subequations}
It is easily verified that $j^{\a(m) \ad(n)}$, $Q^{\a(m+1) \ad(n)}$ and $S^{\a(m) \ad(n+1)}$ define conserved currents satisfying eq. \eqref{ConservedCurrent} for an arbitrary background. 
This is true for $T^{\a(m+1) \ad(n+1)}$ only in the special case where $m = n = 1$.
Let us elaborate on the current \eqref{6.31d} in some more detail. 

In the case of AdS and Minkowski superspaces,  $T^{\a(m+1) \ad(n+1)}$ may always be 
improved,
\bea
\label{CurrentExtension}
\mathbb{T}^{\a(m+1) \ad(n+1)} &:=& T^{\a(m+1) \ad(n+1)} - \frac{2 \ri (m - n)}{m+n+2} \mathfrak{D}^{\a \ad} j^{\a(m) \ad(n)} ~, 
\eea
to give a conserved current,
$\mathfrak{D}_{\b \bd} \mathbb{T}^{\b \a(m) \bd \ad(n)} = 0 $, for arbitrary 
positive integers $m$ and $n$. 
Since the supercurrent $J^{\a(m) \ad(n)}$ is a primary superfield, 
it should be always possible to improve
 \eqref{6.31d} to a conserved current 
in a conformally flat background, $C_{abcd} = 0$.
However, if the background Weyl tensor is non-vanishing, $C_{abcd} \neq 0$, 
it is not possible to improve $T^{\a(m+1) \ad(n+1)} $ to a conserved current provided 
$m>1$ and/or  $n>1$. This conclusion is 
analogous to a recent result of Beccaria and Tseytlin  \cite{BeccariaT} who demonstrated 
that for a conformal scalar field in curved space there is no way to construct a conserved 
traceless symmetric spin-3 current $\cJ^{a bc} $ if the background Weyl tensor is non-vanishing. 

Next, we consider conformal supercurrents of the form $J^{\a(m)}$ \eqref{supercurrent2}. At the component level, it contains two possible candidates for conserved currents:
\begin{subequations}
\bea
j^{\a(m) \ad} &:=& \cDB^{\ad} J^{\a(m)} | ~, \\
T^{\a(m+1) \ad} &:=& [ \cD^{\a} , \cDB^{\ad} ] J^{\a(m)} | ~.
\eea
\end{subequations}
A routine calculation reveals that $j^{\a(m) \ad }$ does indeed constitute a conserved current. 
In the context of AdS and Minkowski superspaces, it is always possible to extend $T^{\a(m) \ad}$ to a conserved current by setting $m=0$ in \eqref{CurrentExtension}, however this fails in the general case.

The final case of interest is that of a scalar conformal supercurrent $J$ \eqref{supercurrent3}. It contains a single current at the component level, 
\bea
T^{\a \ad} := [ \cD^{\a} , \cDB^{\ad} ] J | ~, \qquad \mathfrak{D}_{\a \ad} T^{\a \ad} = 0 ~,
\eea
which is conserved for any curved background.

%%%%%%%%%%%%%%%%%%%%%%%%%%%%%%%%%
%%%%%%%%%%%%%%%%%%%%%%%%%%%%%%%%%

\subsection{Maximally supersymmetric backgrounds}

There exist only  five maximally supersymmetric backgrounds 
in off-shell 4D $\cN=1$  supergravity, as was first demonstrated  
by Festuccia and Seiberg \cite{FS} in the component setting. 
There is a remarkably simple superspace derivation of this classification
\cite{KT-M16,K16} which we review here. Unlike the previous analysis in this section, 
which has relied on the Weyl multiplet gauge, 
this derivation makes use of the gauge condition \eqref{5.7}.

We start by recalling an important  theorem concerning the maximally supersymmetric 
backgrounds \cite{KNT-M,K15Corfu}. 
For any  supergravity theory in $D$ dimensions formulated in superspace, 
all maximally supersymmetric spacetimes correspond 
to those  supergravity backgrounds which 
are characterised by the following properties:
(i) 
all Grassmann-odd components 
of the superspace torsion and curvature tensors vanish; and 
(ii) 
all Grassmann-even components of the torsion and curvature tensors are annihilated 
by the spinor derivatives. 

In the case of 4D $\cN=1$ supergravity, the above theorem means the following:
\begin{subequations}
 \bea
 X_\a &=& 0~,  \label{6.27a}\\
 W_{\a\b\g}&=&0 ~, \label{6.27b} \\
 \cD_\a R =0 ~\implies ~\cD_A R &=&0 ~,  \label{6.27c} \\
 \cD_\a G_{\b\bd} = 0~
\implies~\cD_A G_{\b\bd} &=&0~. \label{6.27d}
\eea
\end{subequations}
Equation \eqref{6.27a} tells us that all maximally supersymmetric backgrounds
 are realised in terms of  the GWZ geometry \cite{GWZ}.
 Equation \eqref{6.27b} tells us that all maximally supersymmetric backgrounds
 are conformally flat.
 Equations \eqref{6.27c} and \eqref{6.27d} restrict  $R$ and $G_{\b\bd} $
 to be covariantly constant. 
Equation \eqref{6.27d} has an integrability condition that follows from 
\bea
0 = \big\{ \bar \cD_\ad , \bar \cD_\bd \big\} G_{\g \gd} 
= 4R \ve_{\gd (\ad} G_{\g \bd)} 
~,
\eea
and therefore we obtain the constraint 
\bea
R G_{\a\ad} =0~. 
\label{6.29}
\eea
There is an alternative way to arrive at this constraint.
Relation  \eqref{6.27d} tells us that $G_{\b\bd}$ satisfies 
the superconformal Killing equation
\eqref{newConformalKilling-b}, and therefore the condition \eqref{3.66} holds. 
Since $G_{\b \bd}$ is covariantly constant, \eqref{3.66} reduces to \eqref{6.29}.

The simplest solution to  \eqref{6.29} is $R=0$ and $G_{\a\ad}=0$, which corresponds to 
Minkowski superspace. Another solution is described by $G_{\a\ad}=0$ and $R=\m \neq 0$, which corresponds to the AdS superspace \eqref{AdSsuperspace}.
The  three remaining superspaces are characterised by 
formally identical anti-commutation relations 
\begin{subequations}\label{RS^3}
\bea
&\{\cD_\a,\cD_\b\}= 0~, \qquad \{\cDB_\ad,\cDB_\bd\}=0~,\qquad
\{\cD_\a,\cDB_\bd\}=-2\ri\cD_{\a\bd}~, 
\\
&{[}\cD_\a,\cD_{\b\bd}{]}=\ri\ve_{\a\b}G^\g{}_{\bd}\cD_\g
~,\qquad
{[}\cDB_\ad,\cD_{\b\bd}{]}=-\ri\ve_{\ad\bd}G_\b{}^\gd\cDB_\gd~,
 \\
&{[}\cD_{\a\ad},\cD_{\b\bd}{]}=
-\ri\ve_{\ad\bd}G_\b{}^\gd\cD_{\a\gd}
+\ri\ve_{\a\b}G^\g{}_\bd\cD_{\g\ad}~,
\eea
\end{subequations}
where  $G_{b} $ is covariantly constant,  $\cD_A G_b = 0$.
The difference between these superspaces is encoded in the Lorentzian type of 
$G_a$.  Since $G^2 = G^a G_a $ is constant, the geometry 
\eqref{RS^3} describes  three different superspaces, 
${\mathbb M}^{4|4}_{T}$, ${\mathbb M}^{4|4}_{S}$ and ${\mathbb M}^{4|4}_{N}$, 
which correspond to the choices $G^2<0$, $G^2>0$ and $G^2=0$, 
respectively.
The Lorentzian manifolds, which are  the bosonic bodies of the superspaces 
${\mathbb M}^{4|4}_{T}$, ${\mathbb M}^{4|4}_{S}$ and ${\mathbb M}^{4|4}_{N}$,
are  ${\mathbb R}\times S^3$, 
${\rm AdS}_3 \times {\mathbb R}$ 
and a pp-wave spacetime, respectively. The latter spacetime is isometric to the so-called Nappi-Witten group  \cite{NappiW}, as shown in \cite{deMF-OS}.

Each superspace \eqref{RS^3} is maximally supersymmetric solution of $R^2$ supergravity \cite{K16}.

%%%%%%%%%%%%%%%%%%%%%%%%%%%%%%%%%%
%%%%%%%%%%%%%%%%%%%%%%%%%%%%%%%%%%%

\section{Conclusion}

To conclude this paper we summarise the main results obtained and 
list several interesting open problems. The main outcomes of this work 
include the following.
\begin{itemize}

\item We described the general structure of (conformal) isometries of supergravity backgrounds within the $\sU(1)$ superspace setting. Using the formalism developed, 
it is trivial to read off the known (conformal) Killing spinor equations for unbroken supersymmetry transformations. What is more important is that our formalism makes it possible to reconstruct, 
starting from a given (conformal) Killing spinor field, 
 a unique (conformal) Killing supervector field which generates the corresponding supersymmetry transformation on $\cM^{4|4}$. 

\item It was shown that  the infinitesimal (conformal) isometry transformations form a closed algebra for any supergravity background. 

\item  We introduced the (conformal) Killing tensor superfields 
$\ell_{\a (m) (\ad (n) }$, where $m$ and $n$ non-negative integers,  $m+n>0$, 
and demonstrated their significance in the following  cases: (i) $m=n$, with the choice $n=1$ corresponding to the (conformal) isometries; 
and (ii) $m-1=n=0$. In particular, we showed that 
extended (conformal) supersymmetry transformations are formulated in terms of the (conformal) Killing spinor superfields $\ell_{\a}$. 
It was proved that the conformal Killing tensor superfields with $m=n$ generate 
all (non-trivial) symmetries of the massless Wess-Zumino operator and form a
superalgebra with respect to the bracket \eqref{4.56}. In the case of conformally flat superspaces this leads to a geometric realisation of the $\cN=1$ conformal higher-spin superalgebra \cite{FL-algebras}.\footnote{All conformal higher-spin superalgebras in four dimensions were classified in \cite{Vasiliev2001}. These results were extended to higher dimensions in \cite{Vasiliev2004}. 
}
 
\item We introduced the conformal  supercurrents 
$J^{\a(m) \ad(n)}$ of arbitrary valence $(m,n)$ in a supergravity background
 and analysed their component structure. 
 
\end{itemize} 

Interesting open problems include the following.
\begin{itemize}
 
 \item We believe that all coefficents of the symmetry operator 
 ${\mathfrak O}^{(n)} $, eq.  \eqref{canonicalForm}, 
 can be expressed in terms of the top component $ \z^{\a(n) \ad(n)} $. 
 We have been able to prove this for the lowest cases $n=1,2$. 
It would be interesting to extend the proof to greater values of $n$.
  
 \item We expect that the  component field defined by  \eqref{6.31d} can be 
  improved to a conserved current $\mathbb{T}^{\a(m+1) \ad(n+1) }$ $( m, n \geq 0  )$, 
   on any conformally flat  bosonic background. A proof of this result would be important.
  Perhaps the best approach to address this problem is to make use of conformal superspace 
  \cite{Butter4DN=1}. 

\item It would be interesting to extend the analysis of section \ref{section4}
to off-shell supergravity backgrounds in diverse dimensions.
In particular, it is an interesting problem to describe the higher symmetries 
of a massless  hypermultiplet in 4D $\cN=2$ conformal supergravity backgrounds.

\item
As an extension of Eastwood's influential work \cite{Eastwood},
there have appeared several publications on higher symmetries of the conformal powers of the Laplacian including \cite{EL,RS,BG2013,LS}.\footnote{The symmetry algebras for higher-derivative equations such as
   $\Box^n$ were actually introduced in the bulk language in \cite{BCIV}.
   See also  \cite{Vasiliev2012,GV13} for further developments.
} 
It would be interesting to carry out a similar analysis
for the $\cN=1$ and $\cN=2$ superconformal extensions of ${\Box}^2$
 proposed in 
\cite{ButterK2013,Butter:2013lta}.

\item General non-conformal deformations of 
the conformal supercurrents $J^{\a(n) \ad(n)}$ and $J^{\a(n+1) \ad(n)}$, eq. \eqref{supercurrent},
 were described in \cite{BHK,HK1,HK2} for the cases of Minkowski and AdS backgrounds.
Various aspects of such non-conformal higher-spin supercurrents in Minkowski superspace 
were studied in \cite{BGK1,KKvU,BGK3}. It would be interesting to study 
consistent non-conformal deformations of other conformal supercurrents introduced
in section \ref{subsection4.22}.

\end{itemize}

%%%%%%%%%%%%%%%%%%%%%%%%%%%%%%%%%%%
%%%%%%%%%%%%%%%%%%%%%%%%%%%%%%%%%%%

\noindent
{\bf Acknowledgements:}\\
Conversations and email correspondence with Misha Vasiliev are gratefully acknowledged.
The work of SK is supported in part by the Australian 
Research Council, project No. DP200101944.
The work of ER is supported by the Hackett Postgraduate Scholarship UWA,
under the Australian Government Research Training Program.

\appendix

\section{Chiral action} \label{AppendixA}

There is an alternative 
way to define the chiral action \eqref{chiralAc}
that follows from the superform approach to 
the construction of supersymmetric invariants \cite{Castellani,Hasler,Ectoplasm,GGKS}.
It is based on the use of
the  following  super four-form 
\bea
\Xi_4[\cL_{\rm c}]&=&
2\ri \bar{E}_\dd \wedge \bar{E}_\gd\wedge  E^b\wedge  E^a(\ts_{ab})^{\gd\dd}
\cL_{\rm c}
+\frac{\ri}{6}\ve_{abcd}\bar{E}_\dd\wedge  E^c\wedge E^b\wedge  E^a (\ts^d)^{\dd\d}\cD_\d\cL_{\rm c}
\non\\
&&
-\frac{1}{96}\ve_{abcd}E^d\wedge E^c\wedge E^b\wedge E^a 
\big(\cD^2-12\bar{R}\big)\cL_{\rm c}~,
\label{A.1}
\eea
which was constructed by Bin\'etruy {\it et al.} 
\cite{Binetruy} and independently by Gates {\it et al.} \cite{GGKS}.\footnote{A
simple derivation of \eqref{A.1}, based on the use of an on-shell vector multiplet, 
was given in \cite{GKT-M}.}
Here we have made use of the superspace vielbein
\bea
E^A = (E^a,E^\a,\bar{E}_\ad) = \rd z^ME_{M}{}^A
~.
\eea
These super one-forms
constitute the dual basis to  
$E_A=(E_a,E_\a,\bar{E}^\ad) = E_A = E_A{}^M  \pa_M$.
The super four-form \eqref{A.1} is closed, 
 \bea
 \rd \, \X_4 [\cL_{\rm c}] =0~.
 \eea
 The chiral action \eqref{chiralAc} can be recast
  as an integral of $\Xi_4[\cL_{\rm c}]$ over a spacetime $\cM^4$,
 \bea
 S_{\rm c} = \int_{\cM^4} \Xi_4[\cL_{\rm c}]~,
 \label{A.3}
 \eea
where $\cM^4$ is the bosonic body of the curved superspace  $\cM^{4|4}$
obtained by switching off  the Grassmann variables. 
 The representation \eqref{A.3} provides the simplest 
 way to reduce the action from superfields to components.

Making use of the super-Weyl transformation laws 
\bea
\d_{\S} E^{a} = - \S E^{a} ~, \qquad 
\d_{\S} E^{\a} = - \frac{1}{2} \S E^{\a} - \frac{\ri}{2}  \cDB_{\bd} E^{b}( \tilde{\s}_{b} )^{\bd \a} ~, 
\eea
it may be shown that 
the super four-form \eqref{A.1} is super-Weyl invariant. 
This result extends the analysis given in  \cite{KT-M17} where the GWZ geometry
was used. In conformal superspace \cite{Butter4DN=1}  the superform \eqref{A.1} was described in \cite{BKN}.

%%%%%%%%%%%%%%%%%%%%%%%%%%%%%%%%%%%%%%%
%%%%%%%%%%%%%%%%%%%%%%%%%%%%%%%%%%%%%%%%

\section{Component reduction}
\label{AppendixB}

To study supergravity-matter theories at the component level, it is necessary to make use of the technique of bar projection. Given a superfield $\X (z)$ defined on $\mathcal{M}^{4|4}$, we define
\bea
\X | ( x ):= \X ( x , \theta , \bar{\theta} ) |_{\theta^{\m} = {\bar \theta}_{\dot{\m}} = 0} ~.
\label{B.1}
\eea
Thus, $\X |$ is a field defined on the background spacetime $\mathcal{M}^{4}$. In the same way, we may define the bar projection of a covariant derivative by bar projecting the connection superfields
\bea
\cD_{A} | \, := E_{A}{}^{M} | \partial_{M} + \frac{1}{2} \O_{A}{}^{bc}| M_{bc} + {\rm i} \F_{A} | \mathbb{A} ~.
\eea
In particular, the bar projected vector covariant derivative takes the form
\bea
\label{barProjcD}
\cD_{a} | \, = \mathfrak{D}_{a} + \frac{1}{2} \psi_{a ,}{}^{\b} \cD_{\b} | + \frac{1}{2} { \bar \psi}_{a , \bd} \cDB^{\bd}| ~,
\eea
where we have introduced both the gravitino $\psi_{a ,}{}^{\b}$ and the charged spacetime covariant derivative
\bea
\label{chargedD}
\mathfrak{D}_{a} = e_{a} + \frac{1}{2} \o_{a}{}^{bc} M_{bc} + \mathrm{i} \varphi_{a} \mathbb{A} ~.
\eea

%%%%%%%%%%%%%%%%%%%%%%%%%%%%%%%%%%%%
%%%%%%%%%%%%%%%%%%%%%%%%%%%%%%%%%%%

\subsection{Wess-Zumino gauge}

By making use of the $\mathcal{K}$ gauge freedom \eqref{gaugeTf}, we are able to fix a Wess-Zumino gauge on the spinor covariant derivatives
\bea
\label{WZgauge}
\cD_{\a} | \, = \d_{\a}{}^{\m} \partial_{\m}, \qquad \cDB^{\ad} | \, = \d^{\ad}{}_{\dot{\m}} \bar{\partial}^{\dot{\m}} ~.
\eea
This gauge leads to the useful identities
\bea
E_{a}{}^{m} | = e_{a}{}^{m} ~, \quad E_{a}{}^{\m} | = \frac{1}{2} \psi_{a}{}^{\b} \d_{\b}{}^{\m} ~, \quad \O_{a}{}^{bc} | = \o_{a}{}^{bc} ~, \quad \Phi_{a} | = \varphi_{a} ~.
\eea
In what follows, we will adopt gauge \eqref{WZgauge}.

Naturally, we are interested in determining the residual gauge transformations which preserve the conditions \eqref{WZgauge}. 
These must satisfy the identity
\bea
\label{preseveWZgauge}
\left( \d_{\mathcal{K}} + \d_{\S} \right) \cD_{\a} | \, = 0 ~.
\eea
The  $\mathcal{K}$ gauge transformations act on the components 
of the connection by the rules:
\begin{subequations}
\bea
\d_{\mathcal{K}} E_{A}{}^{M} & = & \xi^{B} \mathcal{T}_{BA}{}^{C} E_{C}{}^{M} - (\cD_{A} \xi^{B}) E_{B}{}^{M} + K_{A}{}^{B} E_{B}{}^{M} + {\rm i} \r w_{A}{}^{B} E_{B}{}^{M} ~, \\
\d_{\mathcal{K}} \O_{A}{}^{cd} & = & \xi^{B} \mathcal{T}_{BA}{}^{E} \O_{E}{}^{cd} + \xi^{B} \mathcal{R}_{BA}{}^{cd} - (\cD_{A} \xi^{B}) \O_{B}{}^{cd} + K_{A}{}^{B} \O_{B}{}^{cd} - \cD_{A} K^{cd} \non \\
&& + {\rm i} \r w_{A}{}^{B} \O_{B}{}^{cd} ~, \\
\d_{\mathcal{K}} \F_{A} & = & \xi^{B} \mathcal{T}_{BA}{}^{C} \F_{C} + \xi^{B} \mathcal{F}_{BA} - (\cD_{A} \xi^{B}) \F_{B} + K_{A}{}^{B} \F_{B} + {\rm i} \r w_{A}{}^{B} \F_{B} - \cD_{A} \r ~.~~~~
\eea
\end{subequations}
Where we have introduced
\bea
K_{A}{}^{B} =
\left(\begin{array}{ccc}
K_{a}{}^{b}  & 0 &  0\\
 0 & K_{\a}{}^{\b} & 0 \\
 0 & 0 & - {\bar K}^{\ad}{}_{\bd}\\
\end{array}\right)
~, \quad
w_{A}{}^{B} =
\left(\begin{array}{ccc}
0 & 0  &  0\\
0 & - \d_{\a}{}^{\b} & 0\\
0 & 0 & \d^{\ad}{}_{\bd} \\
\end{array}\right) ~.
\eea
By making use of \eqref{superWeylTf}, we extract the super-Weyl transformation laws for the connections:
\begin{subequations}
\bea
\d_{\S} E_{\a}{}^{M} & = & \frac{\S}{2} E_{\a}{}^{M} ~, \\
\d_{\S} \O_{\a}{}^{cd} & = & \frac{\S}{2} \O_{\a}{}^{cd} + 2 ( \sigma^{cd} )_{\a \b} \cD^{\b} \S ~, \\
\d_{\S} \F_{\a} & = & \frac{\S}{2} \F_{\a} - \frac{3 {\rm i}}{2} \cD_{\a} \S ~, \\
\d_{\S} E_{a}{}^{M} & = & \S E_{a}{}^{M} - \frac{\ri}{2} (\tilde{\s}_{a})^{\ad \a} \cD_{\a} \S {\bar E}_{\ad}{}^{M} - \frac{\ri}{2} (\tilde{\s}_{a})^{\ad \a} \cDB_{\ad} \S E_{\a}{}^{M} ~, \\
\d_{\S} \O_{a}{}^{cd} & = & \S \O_{a}{}^{cd} - \frac{\rm i}{2} (\tilde{\s}_{a})^{\ad \a} \cD_{\a} \S { \bar \O}_{\ad}{}^{cd} - \frac{\rm i}{2} (\tilde{\s}_{a})^{\ad \a} { \bar \cD}_{\ad} \S \O_{\a}{}^{cd} + \d_{a}{}^{[c} \cD^{d]} \S  \non \\
&& + \frac{1}{4} \ve_{a}{}^{bcd} (\tilde{\s}_{b})^{\ad \a} [ \cD_{\a} , \cDB_{\ad} ] \S ~, \\
\d_{\S} \F_{a} & = & \S \F_{a} - \frac{ \rm i}{2} (\tilde{\s}_{a})^{\ad \a} \cD_{\a} \S {\bar \F}_{\ad} - \frac{\rm i}{2} (\tilde{\s}_{a})^{\ad \a} \cDB_{\ad} \S \F_{\a} + \frac{3}{8} (\tilde{\s}_{a})^{\ad \a} [ \cD_{\a} ,\cDB_{\ad} ] \S ~.
\eea
\end{subequations}
Thus, \eqref{preseveWZgauge} takes the form
\begin{subequations}
\bea
\cD_{\a} \xi^{\b} | & = & \xi^{C} \mathcal{T}_{C \a}{}^{\b} | + K_{\a}{}^{\b}| - {\rm i} \d_{\a}{}^{\b} \r | + \frac{1}{2} \d_{\a}{}^{\b} \S | ~,\\
\cD_{\a} \bar{ \xi }_{\bd} | & = & \xi^{C} \mathcal{T}_{C \a , \bd}| ~, \\
\cD_{\a} \xi^{b} | & = & \xi^{C} \mathcal{T}_{C \a}{}^{b} | ~, \\
\cD_{\a} K^{cd} | & = & \xi^{B} \mathcal{R}_{B \a}{}^{c d} | - 2 ( \s^{cd} )_{\a}{}^{\b} \cD_{\b} \S | ~, \\
\cD_{\a} \r| & = & \xi^{B} \mathcal{F}_{B \a} | - \frac{3 \rm i}{2} \cD_{\a} \S | ~.
\eea
\end{subequations}
Note that these are equivalent to the bar projection of the conformal Killing conditions \eqref{type2constraints1}. These place severe restrictions on the transformations which preserve this gauge. In particular only the following gauge parameters remain unconstrained
\bea
v^{a} := \xi^{a}| ~, \qquad \e^{\a} := \xi^{\a}| ~, \qquad k_{a b} := K_{a b}| ~, \qquad 
\varrho := \r| ~,
\label{B.11}
\eea
which correspond to general coordinate, local $Q$-supersymmetry, Lorentz and $\sU(1)_{R}$ transformations respectively.

%%%%%%%%%%%%%%%%%%%%%%%%%%
%%%%%%%%%%%%%%%%%%%%%%%%%%

\subsection{Component field strengths}

The (charged) spacetime covariant derivative introduced in \eqref{chargedD} obeys the following commutation relations
\bea
\left[ \mathfrak{D}_{a} , \mathfrak{D}_{b} \right] = T_{a b}{}^{c} \mathfrak{D}_{c} + \frac{1}{2} R_{a b}{}^{c d} M_{c d} + {\rm i} F_{a b} \mathbb{A} ~,
\eea
where $T_{a b}{}^{c}$ is the torsion, $R_{a b}{}^{c d}$ is the Lorentz curvature and $F_{a b}$ is the $\sU(1)_{R}$ field strength. By making use of \eqref{barProjcD} and the bar projection of \eqref{vectorCommutator} it is possible to read off the field strengths.

The simplest field strength to compute is the torsion
\bea
\label{Torsion}
T_{a b c} = - \frac{\rm i}{2} \left( \psi_{a} \s_{c} \bar{\psi}_{b} - \psi_{b} \s_{c} \bar{\psi}_{a} \right) - \ve_{a b c d} G^{d} | ~.
\eea
This result allows us to decompose the Lorentz connection in terms of a torsionless (spin) connection and the torsion
\bea
\label{spinConnection}
\o_{a b c} = \o_{a b c} ( e ) + \frac{1}{2} \left( T_{abc} - T_{bca} + T_{cab} \right) ~.
\eea

It is also convenient to introduce the gravitino field strength
\bea
\Psi_{a b}{}^{\g} := \mathfrak{D}_{a} \psi_{b , }{}^{\g} - \mathfrak{D}_{b} \psi_{a , }{}^{\g} - T_{ab}{}^{c} \psi_{c , }{}^{\g} ~,
\eea
which can be computed to be
\bea
\label{GravitinoFS}
\Psi_{a b}{}^{\g} & = & - \ri \psi_{[ a}{}^{\a} ( \s_{b ]})_{\a \ad} G^{\ad \g} | - \ri {\bar \psi}_{[a , \ad} ( {\tilde \s}_{b]})^{\ad \g} R | \non \\
&& - \frac{\ri}{2} ( {\tilde \s}_{a b})^{\ad \bd} \psi_{\a \ad , }{}^{\g} G^{\a}{}_{\bd} | - \frac{\rm i}{2} (\s_{a b})^{\a \b} \psi_{ \a \dot{\lambda} , }{}^{\g} G_{\b}{}^{\dot{\lambda}} | + (\s_{a b})^{\a \b} W_{\a \b}{}^{\g}| \non \\
&& + \frac{1}{6} (\s_{ab})^{\g \a} X_{\a}| + \frac{1}{2} (\s_{a b})^{\g \a} \cD_{\a} R | + \frac{1}{2} ( \tilde{\s}_{a b} )^{\ad \bd} \cDB_{( \ad} G^{\g}{}_{\bd)}| ~.
\eea
Next, the $\sU(1)_R$ field strength is given by
\bea
\label{U1FS}
F_{ab} = \frac{1}{8} ( \s_{ab})^{\a \b} \left( {\rm i} \cD_{(\a} X_{\b)} | + {\bar \psi}_{(\a \ad , }{}^{\ad} X_{\b)} | - \psi_{(\a \ad , \b)} {\bar X}^{\ad} |\right) + \rm{c.c.} ~.
\eea
Finally, we compute the Lorentz curvature
\bea
\label{LorentzCurvature}
R_{abcd} & = & \frac{1}{2} \left( {\rm i} \eta_{d e} \eta_{c [a} - {\rm i} \eta_{ce} \eta_{d [a} + \ve_{c d e [a } \right) \psi_{b] , }{}^{\a} ( \s^{e} )_{\a \ad} \cDB^{\ad} {\bar R}| - \frac{ \rm i}{2} (\tilde{\s}_{[ a})^{\ad \a} \psi_{b ] , \a} ( \s_{c d} )^{\g \d} \cD_{\g} G_{\d \ad} | \non \\
&& + {\rm i} ( \tilde{ \s }_{ [ a})^{\ad \a} { \bar \psi}_{b] , \ad} ( \s_{cd} )^{\b \g}  W_{\a \b \g} | + \frac{1}{12} \left( {\rm i} \eta_{d e} \eta_{c [a} - {\rm i} \eta_{ce} \eta_{d [a} + \ve_{c d e [a } \right) \bar{\psi}_{b] , \ad} (\tilde{\s}^{e})^{\ad \ad} X_{\a} | \non \\
&& + \frac{1}{16} \left( \eta_{a c} \eta_{b d} - \eta_{a d} \eta_{b c} + {\rm i} \ve_{abcd} \right) \left( \cDB^{2} \bar{R} | - 8 R | \bar{R} | \right) + \psi_{a} \sigma_{cd} \psi_{b} \bar{R} | \non \\
&& - \frac{1}{4} (\tilde{\s}_{a b})^{\ad \bd} (\s_{c d})^{\a \b} \cDB_{\ad} \cD_{\a} G_{\b \bd} | + \frac{1}{2} ( \s_{a b})^{\a \b} (\s_{c d})^{\g \d} \cD_{\a} W_{\b \g \d} | \non \\
&& + \frac{1}{48} \left( \eta_{d e} \eta_{c [a} - \eta_{c e} \eta_{d [a} + {\rm i} \ve_{c d e [a} \right) ( \s_{b ]} \tilde{\s}^{e} )^{\a \b} \cD_{\a} X_{\b} | \, + \, \rm{c.c.} ~,
\eea
When working at the component level, it is often necessary to understand the relationship between the irreducible components of these field strengths and the component structure of the torsion superfields and their derivatives.

We begin with an analysis of the gravitino field strength $\Psi_{ab}{}^{\g}$, which yields
\begin{subequations}
\label{PsiComponents}
\bea
\cD_{\a} R | + \frac{1}{3} X_{\a} | & = & - \frac{4}{3} \Psi_{\a}{}^{\b}{}_{,\b} + 6 \ri \psi^{\b \ad}{}_{, ( \a} G_{\b ) \ad} | - 9 \ri  {\bar \psi}_{\a}{}^{\ad}{}_{, \ad} R | ~, \\
W_{\a \b \g} | & = & \Psi_{ ( \a \b , \g ) } - \ri \psi_{ ( \a }{}^{\ad}{}_{, \b} G_{\g ) \ad} | ~, \\
\cDB_{ ( \ad } G_{\b \bd )} | & = & - 2 \Psi_{ \ad \bd , \b } - \ri \psi_{\b ( \ad ,}{}^{\a} G_{\a \bd )} | 
+ \ri {\bar \psi}_{\b ( \ad , \bd )} R |  ~.
\eea
\end{subequations}
Moving on to the $\sU(1)_{R}$ field strength $F_{ab}$, we have a single irreducible component
\bea
\label{U1Components}
\cD_{ ( \a } X_{\b )} | = - 8 \ri F_{\a \b} + 8 \ri {\bar \psi}_{(\a \ad ,}{}^{\ad} X_{\b )} | - 8 \ri \psi_{(\a \ad , \b )} \bar{ X }^{\ad} | ~.
\eea
The remaining relations arise from the Lorentz curvature $R_{abcd}$
\begin{subequations}
\label{LorentzComponents}
\bea
\cD^{2} R | & = & \frac{2}{3} \left( R( e , \psi) - \frac{\rm i}{2} \ve_{abcd} R^{abcd} \right) + 2 \ri {\bar \psi}^{\a \ad}{}_{, \ad} \cD_{\a} R | + \frac{ 2 \ri}{3} {\bar \psi}^{\a \ad , \bd} \cDB_{(\ad} G_{\a \bd)} | \non \\
&& - 3 \ri \psi^{\a \ad}{}_{,\a} {\bar X}_{\ad} | + \frac{2}{3} {\bar \psi}^{\a \ad}{}_{, ( \ad} {\bar \psi}_{\a}{}^{\bd}{}_{, \bd )} R | + 8 R | {\bar R }| - \frac{1}{3} \cD X | ~, \\
\cDB_{ ( \ad} \cD_{ ( \a} G_{\b ) \bd)} | & = & 2 E_{\a \b , \ad \bd} + \ri \psi^{\g}{}_{( \ad , \g} \cD_{( \a } G_{\b ) \bd )} | - 2 \ri {\bar \psi}^{\g}{}_{(\ad , \bd )} W_{\a \b \g} | + \frac{\ri}{3} {\bar \psi}_{(\a ( \ad , \bd )} X_{\b )}| \non \\
&& - \psi^{\g}{}_{(\ad , ( \a} \psi_{\g \bd) , \b )} {\bar R} | + 2 \ri \psi_{(\a ( \ad , \b )} \cDB_{\bd)} {\bar R}| ~, \\
\cD_{(\a} W_{\b \g \d )} | & = & C_{\a \b \g \d} + \ri \psi_{(\a \ad}{}_{, \b} {\bar \Psi}_{\g \d) ,}{}^{\ad} + \ri {\bar\psi}_{(\a \ad ,}{}^{ \ad} \Psi_{\b \g \d )} - {\bar \psi}_{(\a \ad ,}{}^{\ad} \psi_{\b \bd ,}{}_{ \g} G_{\d )}{}^{\bd} | \non \\ && - 2 {\bar \psi}_{(\a \ad , \bd} \psi_{\b}{}^{\ad}{}_{, \g} G_{\d )}{}^{\bd} | ~.
\eea
\end{subequations}
Where we have defined
\bea
R(e,\psi) = \eta^{a c} \eta^{b d} R_{abcd} ~, \quad E_{\a \b , \ad \bd} = \frac{1}{2} R^{\g}{}_{( \ad , \g \bd ) , \a \b} ~, \quad C_{\a \b \g \d} = \frac{1}{2} R_{ (\a }{}^{\ad}{}_{,\b \ad , \g \d )} ~.
\eea
It is well known that $C_{\a \b \g \d}$ is the spinor form of the anti-self-dual part of the usual Weyl tensor and as a result $W_{\a \b \g}$ is often referred to as the `super Weyl tensor'. Similarly, $E_{\a \b , \ad \bd}$ coincides with the traceless component of the Ricci tensor and so we say that $G_{\a \ad}$ is its supersymmetric extension.

%%%%%%%%%%%%%%%%%%%%%%%%%%%
%%%%%%%%%%%%%%%%%%%%%%%%%%%

\section{The Weyl multiplet gauge}\label{Weyl_multiplet}

It is often advantageous to adopt a gauge which partially fixes the super-Weyl freedom \eqref{superWeylTf} in exchange for gauging several (component) fields to zero. We recall that in $\sU(1)$ superspace this freedom is parametrised by a real scalar superfield $\S$ \eqref{superWeylTf}, thus it contains six independent component fields in its multiplet.

The component fields $\s := \S |$ and $\eta_{\a} := \cD_{\a} \S |$ parametrise Weyl and S-supersymmetry transformations, respectively. Recalling equations \eqref{superWeylTfTorsions}, we observe that by making use of our freedom in the $\cD^{2} \S |$ and $\left[ \cD_{\a} , \cDB_{\ad} \right] \S |$ component fields, it is possible to adopt a gauge where $R | = \bar{R} | = 0$ and $G_{\a \ad} | = 0$. Further, by a routine calculation one can derive
\begin{subequations}
\bea
\d_{\S} \left( \cD_{\a} R \right) & = & \frac{3}{2} \S \cD_{\a} R + 4 \cD_{\a} \S R + \frac{1}{2} \cD_{\a} \cDB^{2} \S ~, \\
\d_{\S} \left( \cD^{2} R \right) & = & 2 \S \cD^{2} R + 4 \cD^{2} \S R + 4 \cD^{\a} \S \cD_{\a} R + \frac{1}{2} \cD^{2} \cDB^{2} \S ~, \\
\d_{\S} \left( \cD^{2} R + \cDB^{2} {\bar R} \right) & = & 2 \S \left( \cD^{2} R + \cDB^{2} {\bar R} \right) + 4 \cDB^{2} \S \bar{R} + 4 \cD^{2} \S R \non \\
&&+ 4 \cD^{\a} \S \cD_{\a} R + 4 \cDB_{\ad} \S \cDB^{\ad} {\bar R} + \frac{1}{2} \{ \cD^{2} , \cDB^{2} \} \S ~.
\eea
\end{subequations}

By making use of the $R | = 0$ gauge condition, it is possible to use the freedom in $\cD_{\a} \cDB^{2} \S |$ to fix $\cD_{\a} R | = 0$. Finally, we can further extend the gauge by using $\{ \cD^{2} , \cDB^{2} \} \S |$ to set $\cD^{2} R| + \cDB^{2} {\bar R}| = 0$. This completes our gauge fixing procedure. 

We must also determine the residual combined gauge \eqref{tensorgaugeTf} and super-Weyl transformations which preserve this gauge.
A routine computation leads to the conditions
\begin{subequations}
\label{WMresidual}
\bea
\cD^{2} \S | & = & ( \xi^{\a \ad} \cD_{\a \ad} \bar{R} ) | ~, \\
\left[ \cD_{\a} , \cDB_{\ad} \right] \S | & = & - ( \xi^{B} \cD_{B} G_{\a \ad} ) | ~, \\
\cD_{\a} \cDB^{2} \S | & = & - 2 ( \xi^{B} \cD_{B} \cD_{\a} R  ) |  ~, \\
\{ \cD^{2} , \cDB^{2} \} \S | & = & ( \xi^{\a \ad} \cD_{\a \ad} ( \cD^{2} R + \cDB^{2} \bar{R} ) ) |  
+ 8 \ri ( \bar{\xi}^{\ad} \cD_{\a \ad} \cD^{\a} R ) | \non \\
&& + 8 \ri ( \xi^{\a} \cD_{\a \ad} \cDB^{\ad} \bar{R} ) | ~.
\eea
\end{subequations}

In summary, adopting the Weyl multiplet gauge has allowed us to fix
\bea
\label{WMGauge}
R | = 0 ~, \quad G_{\a \ad} | = 0 ~, \quad \cD_{\a} R | = 0 ~, \quad \cD^{2} R | + \cDB^{2} {\bar R}| = 0 ~,
\eea
while retaining unbroken Weyl $ \s $ and S-supersymmetry transformations $ \eta_{\a}$ .

Now, we return to our discussion of the field strengths \eqref{Torsion}, \eqref{GravitinoFS} , \eqref{U1FS} and \eqref{LorentzCurvature}. By imposing \eqref{WMGauge}, we find that these take the simplified form
\begin{subequations}
\bea
T_{abc} & = & - \frac{ \rm i}{2} \left( \psi_{a} \s_{c} \bar{\psi}_{b} - \psi_{b} \s_{c} \bar{\psi}_{a} \right) ~, \\
\Psi_{a b ,}{}^{\g} & = & ( \s_{a b} )^{\a \b} W_{\a \b}{}^{\g} | + \frac{1}{6} ( \s_{a b} )^{\g \a} X_{\a} | + \frac{1}{2} ( \tilde{\s}_{a b} )^{\ad \bd} \cDB_{\ad} G^{\g}{}_{\bd} | \label{C.3b} ~, \\
R_{abcd} & = & - \frac{ \rm i}{2} (\tilde{\s}_{[ a})^{\ad \a} \psi_{b ] , \a} ( \s_{c d} )^{\g \d} \cD_{\g} G_{\d \ad} | + {\rm i} ( \tilde{ \s }_{ [ a})^{\ad \a} { \bar \psi}_{b] , \ad} ( \s_{cd} )^{\b \g}  W_{\a \b \g} | \non \\
&& + \frac{1}{12} \left( {\rm i} \eta_{d e} \eta_{c [a} - {\rm i} \eta_{ce} \eta_{d [a} + \ve_{c d e [a } \right) \bar{\psi}_{b] , \ad } (\tilde{\s}^{e})^{\ad \a} X_{\a} | + \frac{\rm i}{16} \ve_{abcd} \cDB^{2} \bar{R} | \non \\
&& - \frac{1}{4} (\tilde{\s}_{a b})^{\ad \bd} (\s_{c d})^{\a \b} \cDB_{\ad} \cD_{\a} G_{\b \bd} | + \frac{1}{2} ( \s_{a b})^{\a \b} (\s_{c d})^{\g \d} \cD_{\a} W_{\b \g \d} | \non \\
&& + \frac{1}{48} \left( \eta_{d e} \eta_{c [a} - \eta_{c e} \eta_{d [a} + {\rm i} \ve_{c d e [a} \right) ( \s_{b ]} \tilde{\s}^{e} )^{\a \b} \cD_{\a} X_{\b} | \, + \, \rm{c.c.} ~, \\
F_{ab} & = & \frac{1}{8} ( \s_{ab})^{\a \b} \left( {\rm i} \cD_{(\a} X_{\b)} | + {\bar \psi}_{\a \ad , }{}^{\ad} X_{\b} | - \psi_{\a \gd, \b} {\bar X}^{\gd} |\right) + \rm{c.c.} ~.
\eea
\end{subequations}
Another advantageous property of this choice of gauge is that the relations \eqref{PsiComponents} , \eqref{U1Components} and \eqref{LorentzComponents} are greatly simplified. We read off
\begin{subequations}
\bea
\mkern-18mu  X_{\a} | & = & - 4 \Psi_{\a}{}^{\b}{}_{,\b} ~, \quad W_{\a \b \g} | = \Psi_{ ( \a \b , \g) } ~, \quad \cDB_{( \ad} G_{\b \bd )} | = -2 \Psi_{\ad \bd ,\b} \label{C.4a} ~, \\
\mkern-18mu \cD_{(\a} X_{\b)} | & = &  - 8 {\rm i} F_{\a \b}  + 32 {\rm i} {\bar \psi}_{ ( \a }{}^{\ad}{}_{,\ad} \Psi_{\b) \g ,}{}^{\g} - 32 {\rm i} \psi_{ ( \a }{}^{\ad}{}_{ , \b )} { \bar \Psi}_{\ad}{}^{\bd}{}_{, \bd} ~, \\
\cD X | & = & 2 R ( e , \psi ) + \left( 18 \ri \psi^{\a \ad}{}_{, \a} {\bar \Psi}_{\ad}{}^{\bd}{}_{, \bd} - 2 \ri \psi^{\a \ad , \b} {\bar \Psi}_{\a \b , \ad} + \text{c.c.} \right) ~, \\
\mkern-18mu \cD^{2} R | & = & - \frac{\rm i}{3} \ve^{abcd} R_{abcd} + \ri \left( 6 \psi^{\a \ad}{}_{, \a} {\bar \Psi}_{\ad}{}^{\bd}{}_{, \bd} + \frac{2}{3} \psi^{\a \ad , \b} {\bar \Psi}_{\a \b , \ad} + \rm{c.c.} \right) ~, \\
\mkern-18mu \cD_{( \ad } \cD_{ ( \a } G_{\b ) \bd )} | & = & 2 E_{\a \b , \ad \bd} - 2 \ri \psi^{\g}{}_{( \ad , \g} {\bar \Psi}_{\a \b , \bd )} - 2 {\rm i} {\bar \psi}^{\g}{}_{ ( \ad , \bd ) } \Psi_{ ( \a \b , \g )} - \frac{4}{3} \ri {\bar \psi}_{( \a ( \ad , \b )} \Psi_{\b)}{}^{\g}{}_{\g} ~, \\
\mkern-18mu \cD_{ (\a } W_{\b \g \d )} | & = & C_{\a \b \g \d} + {\rm i} {\bar \psi}_{(\a \ad ,}{}^{\ad} \Psi_{\b \g , \d )} + {\rm i} \psi_{(\a \ad , \b} {\bar \Psi}_{\g \d ) ,}{}^{\ad} ~.
\eea
\end{subequations}

When combined with the algebra \eqref{algebra} and Bianchi identites \eqref{Bianchi}, these relations allow us to express all component fields of the torsion superfields in terms of the (component) field strengths. The main implication of this is that the only remaining independent component fields are the spacetime vielbein $e_{m}{}^{a}$, the gravitino $\psi_{m}{}^{\a}$ (and its conjugate) and the $\sU(1)_{R}$ gauge field $\varphi_{m}$, which are known to comprise the Weyl multiplet.

The approach described in this appendix is analogous 
to the one used for 3D $\cN=2$ supergravity \cite{KLRST-M}.

\begin{footnotesize}

\end{footnotesize}

%%%%%%%%%%%%%%%%%%%%%%%%%%%%%%%%%%%%
%%%%%%%%%%%%%%%%%%%%%%%%%%%%%%%%%%%%

\end{document}